%% file: main.tex
\documentclass[11pt,a4paper]{article}
\pdfoutput=1
\usepackage{jinstpub}
\usepackage[english]{babel}
\usepackage{upgreek}
\usepackage{xspace}
\usepackage{graphicx}
\usepackage[capitalize]{cleveref}
\usepackage{enumitem}

\graphicspath{{figs/}}

\title{Reconstruction of Events Recorded with the Surface Detector of the Pierre Auger Observatory}

\author{\includegraphics[height=30mm]{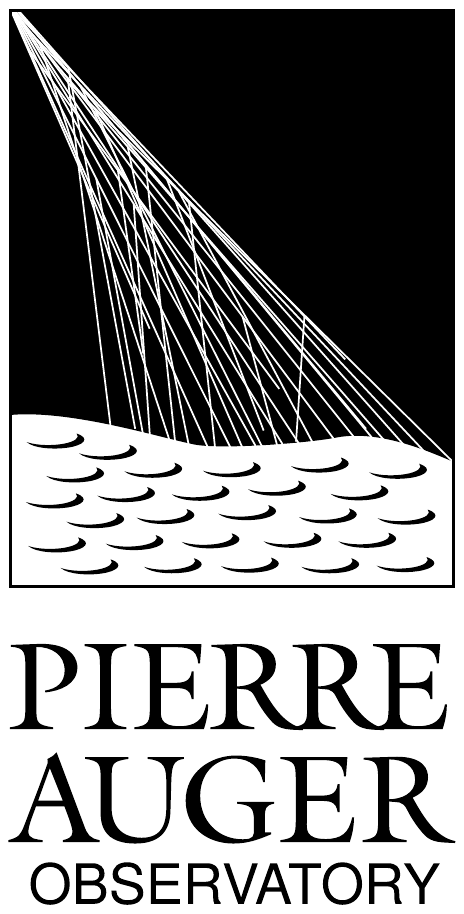}\\[3mm]The Pierre Auger Collaboration}
\affiliation{Av.\ San Mart\'{\i}n Norte 306, 5613 Malarg\"ue, Mendoza, Argentina}

\emailAdd{auger\_spokespersons@fnal.gov}

\abstract{
Cosmic rays arriving at Earth collide with the upper parts of the atmosphere, thereby inducing extensive air showers.
When secondary particles from the cascade arrive at the ground, they are measured by surface detector arrays.
We describe the methods applied to the measurements of the surface detector of the Pierre Auger Observatory to reconstruct events with zenith angles less than $60^\circ$ using the timing and signal information recorded using the water-Cherenkov detector stations.
In addition, we assess the accuracy of these methods in reconstructing the arrival directions of the primary cosmic ray particles and the sizes of the induced showers.
}

\keywords{Large detector systems for particle and astroparticle physics, Data processing methods, Large detector-systems performance, Performance of High Energy Physics Detectors}

\arxivnumber{2007.09035}

\dedicated{\flushleft\textnormal{\textsc{%
Published in JINST as \href{http://dx.doi.org/10.1088/1748-0221/15/10/P10021}{DOI:10.1088/1748-0221/15/10/P10021}
\\
Report number: FERMILAB-PUB-20-487-AD-E-TD
}}}

\begin{document}

\maketitle
\flushbottom

\section{Introduction}

The intensity of cosmic rays with energies above ${\sim}10^{14}$\,eV is only a few particles per square meter per day and thus too low to allow for direct measurement with satisfactory statistical precision.
The phenomenon of extensive air-showers must therefore be exploited to study cosmic rays at higher energies.
In extensive air-showers, the interaction of a primary cosmic-ray with Earth's atmosphere induces a cascade of particles, where the particles reaching the ground may be sampled by large arrays of detectors.

The higher the energies of primary cosmic rays, the lower is their arrival rate such that increasingly larger detection areas are needed to counteract the lower intensity.
Meanwhile, the higher the primary energy, the larger are the dimensions of the resulting air shower such that an increased spacing between detectors is sufficient to ensure shower detection.
For example, the footprint at sea level of a shower generated by a primary cosmic-ray with an energy of $10^{19}$\,eV is ${\sim}10$\,km$^2$, and it has been found empirically over the years that having several detectors inside such an area is sufficient to reconstruct the properties of such showers.
Above $10^{19}$\,eV, the intensity is only ${\sim}1$ per km$^2$ per year, which therefore requires huge collecting areas to gather adequate statistics.

A shower can be thought of as a thin, radially extended, and slightly curved disk of particles propagating longitudinally at the speed of light along the initial direction of the primary cosmic-ray.
At every stage of development, the particle density in the disk is largest at the center, or core, of the shower and decreases with radial distance from the shower axis.
At sea level, about 90\% of the energy flow lies within less than 100\,m of the shower axis; however, gathering a reasonable number of events here would require a prohibitively dense array and a large dynamic range of the particle detectors.
A sparse array of detectors at the ground can only sample the disk at a fixed stage of shower development and only beyond a certain distance from the core. 
The arrival direction and a proxy of the shower size can nevertheless be reconstructed and interpreted in terms of the properties of the initial cosmic ray without the need for deeper insight into the complex physics of extensive air-showers, where the latter quantity can be used to infer the energy of the primary particle.

The arrival direction of the primary cosmic-ray can be determined from the relative times of arrival of the shower disc at the dispersed detectors.
This was first demonstrated in the 1950s by the researchers in the MIT group~\cite{Bassi:1953jja}, who used their newly-developed liquid scintillation detectors to show, for the first time, that the shower disc was sufficiently thin to enable determination of the direction of the shower axis to within a few degrees.
Since then, the detectors of choice have either been scintillators (liquid or plastic) or water-Cherenkov detectors, the use of which was pioneered a few years later~\cite{Porter1958}.

The number of particles in the shower (also referred to as the shower size) at the observation level is related to the primary energy, but because of fluctuations in shower development, the association is not a simple one.
A lower-energy primary particle that interacts deep in the atmosphere can create a cascade which has the same size at the ground as the one produced by a more energetic particle interacting higher in the atmosphere.
Measuring the shower size at the depth of the shower maximum (or better, the integral of the deposited energy along the shower profile) reduces the impact of such fluctuations; however, observations at the shower maximum have only been possible at high altitudes and for relatively low-energy events (see e.g.~\cite{Hersil:1961zz}), or by the observation of the profile with fluorescence detectors~\cite{Bergeson:1977nw,Baltrusaitis:1985mx}, albeit with an order of magnitude lower duty cycle.

When considering depths lower in the atmosphere than the position of the shower maximum, and when using a widely-spaced array of detectors, the situation is more complex.
Determining the fall-off of the signal as a function of distance from the axis is difficult when the spacing of the detectors is large, and it cannot be determined with adequate precision on an event-by-event basis.
A technique to circumvent this problem was introduced by Hillas~\cite{Hillas:1969zzb}, who proposed using the signal at a distance appropriate for the separation of the detectors at which the uncertainty in the determination of this signal is least susceptible to the lack of knowledge of the fall-off of the signal with distance.

The Pierre Auger Observatory~\cite{ThePierreAuger:2015rma} is the world's largest extensive air-shower detector and makes use of these techniques to study cosmic rays at the highest energies.
The hybrid design of the Observatory consists of 27 telescopes of the fluorescence detector (FD)~\cite{Abraham:2009pm} distributed between four sites overlooking the huge array of the surface detector (SD).
The SD, described in the following section, provides the bulk of the events recorded by the Observatory, which are principally used to measure the distribution of arrival directions (e.g.~\cite{Aab:2017tyv}) and the energy spectrum (e.g.~\cite{spectrum_prd_2020}) of the highest-energy cosmic rays.
The reconstruction of these SD events is the focus of this paper.

In this document, we provide a detailed description of the methods used to reconstruct the arrival direction and size of an extensive air-shower from the magnitudes and temporal distributions of signals detected in each SD station.
A critical evaluation of the accuracy of the reconstruction procedure is also an essential part of this work.
The methodology described here is used only for events with a zenith angle $\theta<60^\circ$, as the reconstruction of more inclined showers requires different methods due to the substantial asymmetry induced in the spatial distribution of the shower particles by the geomagnetic field and due to geometrical effects~\cite{Aab:2014gua}.

Within the Pierre Auger Collaboration, two independent frameworks have been developed for the reconstruction of these events with the corresponding releases of reconstructed data dubbed \emph{Herald} and \emph{Observer}~\cite{Argiro:2007qg}.
The existence of these independent analyses facilitates easy access to cross-checks, validations, and verification of systematic effects.
The two reconstructions are based on a similar strategy; however, they include slight differences in the description of the temporal and lateral structure of showers.
Such differences are highlighted where relevant.

The paper is organized as follows.
In \cref{s:sd}, we give a brief introduction to the features of the SD relevant for the reconstruction procedures.
In \cref{s:stationrec}, we describe the treatment of the data from individual detectors, including the derivation of the start-times and the magnitudes of the recorded signals.
In \cref{s:evebuilding}, we describe how shower events are built prior to the reconstruction procedure.
In turn, this reconstruction procedure is described in detail in \cref{s:everec}, beginning with the methods used to reconstruct the geometry of the shower (arrival direction of the shower and the impact point of the core) and ending with the estimation of the shower size.
In \cref{s:angularuncer} and in \cref{s:s1000uncer}, the uncertainties associated with the reconstructed parameters are discussed.
In \cref{s:energy_estimator}, we finalize the description of the reconstruction procedures by briefly outlining the assignment of energy.

\section{The surface detector}
\label{s:sd}

The SD consists of more than 1600 water-Cherenkov stations arranged on a isometric triangular grid covering 3000\,km$^2$.
Station elevations range between 1300 and 1600\,m above sea level, and each station is separated from its six nearest neighbors by approximately 1500\,m.
For example, a primary cosmic ray with an energy of $10^{19}$\,eV will trigger at least 5, but on average ${\sim}8$, stations.
Within the boundary of the 1500\,m array, there are two denser regions: one extending over 28\,km$^2$ within which the stations are spaced 750\,m apart from one another, and the other covering 2\,km$^2$ with 433\,m spacing.
For events recorded in these parts of the SD, specific reconstructions are developed based on the reconstruction given here, but appropriately scaled to a smaller spacing.
A detailed description of the SD can be found in~\cite{ThePierreAuger:2015rma}, and only the main features relevant to the work presented in this paper are summarized here.

Each station is a cylindrical container with a base area of 10\,m$^2$. These containers are filled to a depth of 1.2\,m with highly purified water enclosed by a diffusively-reflective liner.
The volume of water is viewed from above by three 9-inch photomultiplier tubes (PMTs), which detect Cherenkov light emitted during the passage of charged particles.
Each PMT provides two signals which are tagged with the GPS time stamps to an absolute time accuracy of ${\sim}12$\,ns~\cite{Allison:2005vj} and are digitized by 40\,MHz, 10-bit Flash Analog-to-Digital Converters (FADCs).
The \emph{low-gain} signal is taken directly from the anode of the PMT, while the \emph{high-gain} signal is provided by the last dynode and amplified to be nominally 32 times larger than the low gain, thus enlarging the total dynamic range to span more than three orders of magnitude in integrated signal.

The data acquisition is governed by a hierarchical system of triggers implemented in hardware and software.
The first two levels, T1 and T2, are formed locally by each detector station~\cite{Abraham:2010zz}.
The set of stations passing the T2-level criteria are continuously examined remotely for temporal and spatial coincidences in at least three stations.
Upon identification of such a coincidence, an array-level trigger T3 is issued by the central data acquisition system (CDAS).
After the T3 trigger is raised, FADC data from the T1 and T2-triggered stations, as well as calibration and monitoring data, are transmitted to CDAS by means of local-area-network radio-communication.

\begin{figure}[t]
\def\figh{0.5}
\centering
\includegraphics[height=\figh\columnwidth]{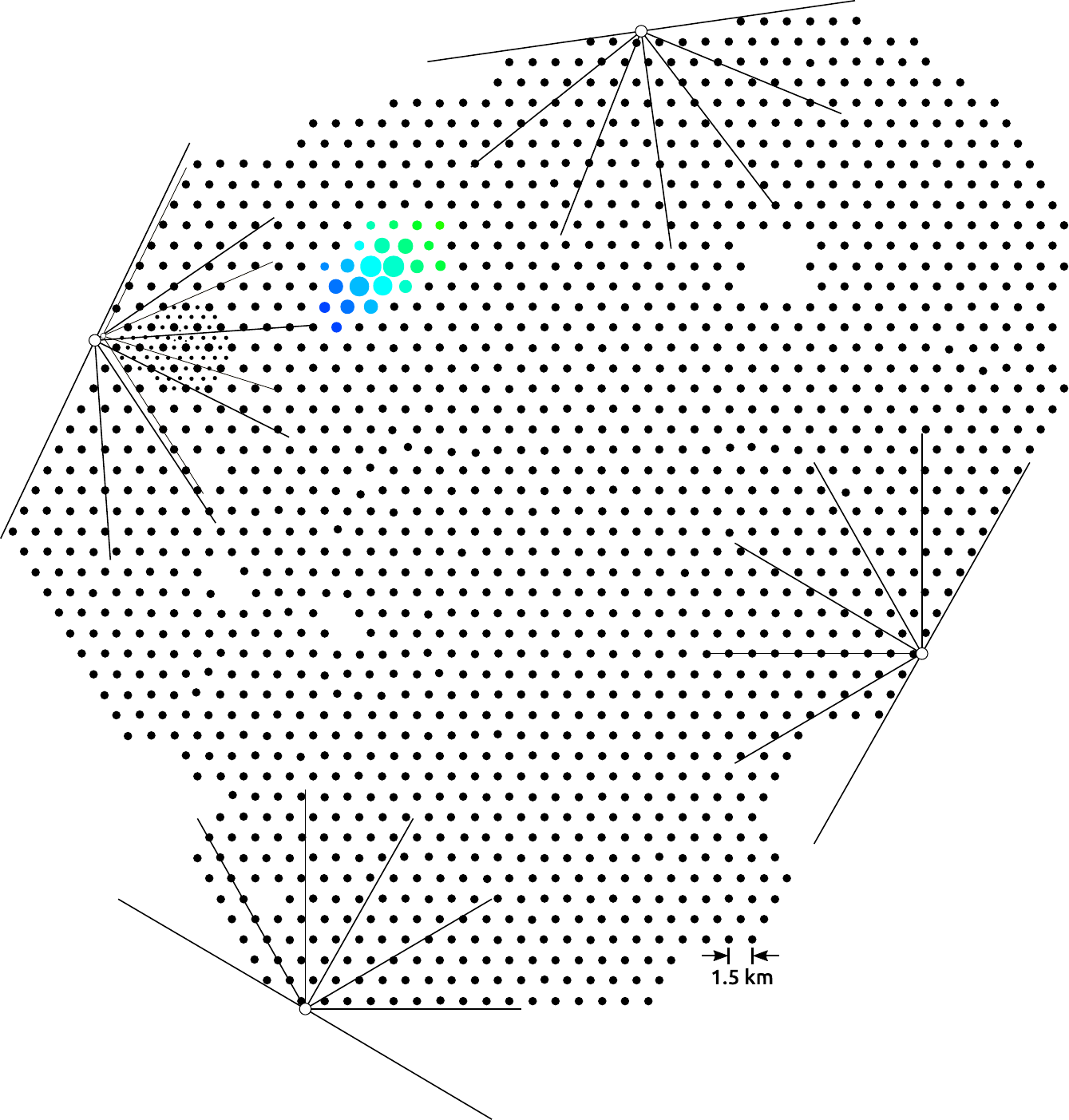}\hfill
\includegraphics[height=\figh\columnwidth]{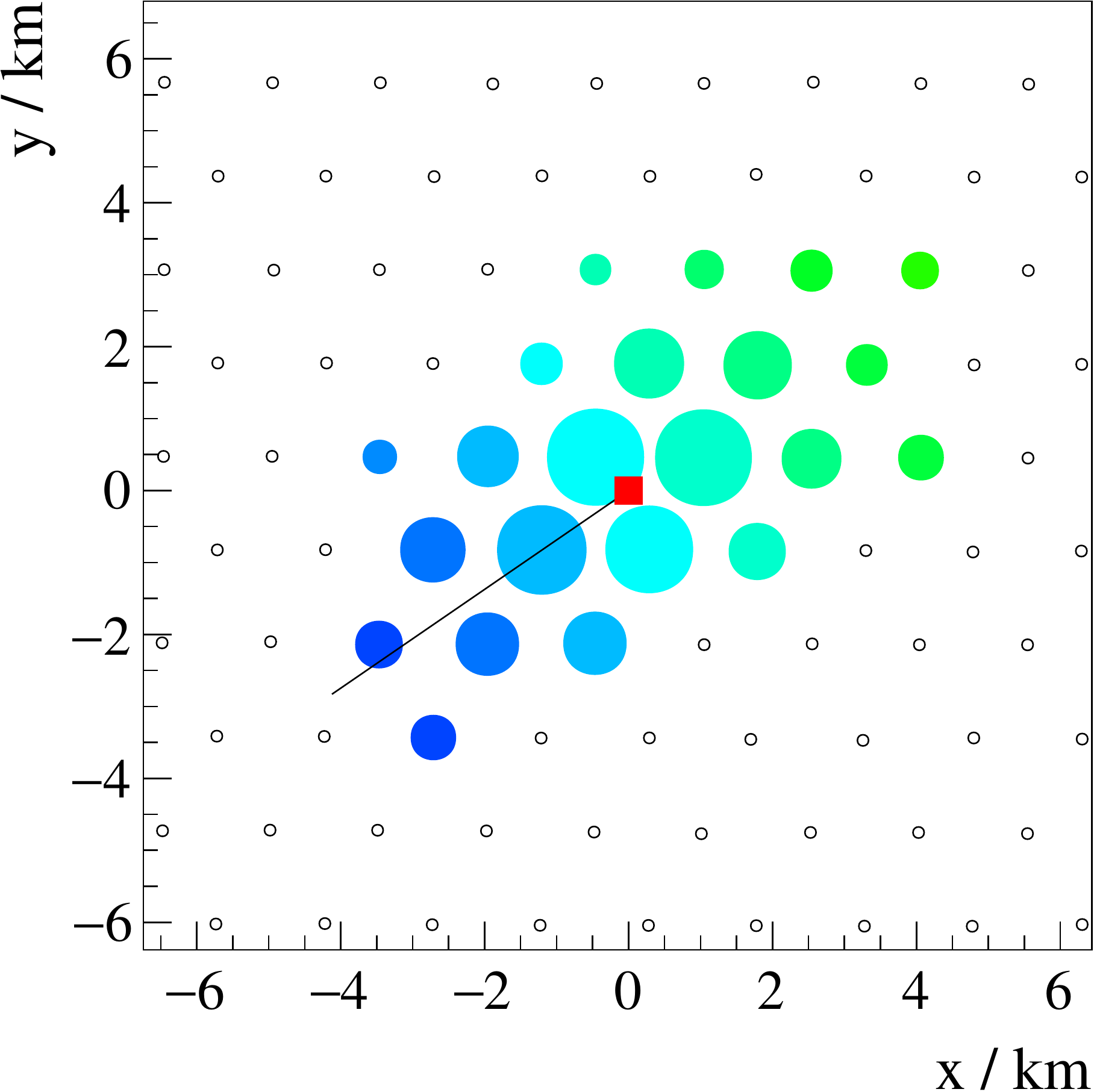}
\caption{Example of a shower event detected by the SD, which is reconstructed with $\theta=55.2^\circ$ and an energy of 38.7\,EeV.
\emph{Left:} Schematic depiction of the surface detector array in which each dot represents a surface detector station. The colored stations correspond to those participating in this example event.
Rays correspond to the field of view of the telescopes at the four FD sites.
\emph{Right:} A zoomed-in, top-down view of the example event falling within the array. The projection of the shower axis on the ground is represented by the black line, ending at the impact-point of the shower-core (red square).
The SD stations are colored according to their trigger time (blue is early, green is late) with their area proportional to the logarithm of the signal.}
\label{f:event}
\end{figure}

Additional higher-level triggers are then implemented off-line to select shower events (physics trigger T4) and to ensure the shower is well contained in the array.
With respect to the latter, we employ a fiducial trigger $N$T5, where $N$ is the number of functioning stations surrounding the station with the highest signal at the time of the event.
In this work we consider only the most stringent condition, the 6T5, where all six neighboring stations must have been present and functioning.
This ensures an accurate reconstruction of the impact point.
For the data set used in this work, 6T5 events\footnote{~This requirement not only ensures adequate sampling of the shower but also allows evaluation of the aperture of the SD in a purely geometrical manner in the regime where the array is fully efficient~\cite{Abraham:2010zz}.} from the period of 1~January 2004 to 31~December 2018 have been selected, resulting in a set of more than 400\,000 events.
However, the results discussed in this paper also apply to the events with a less conservative trigger 5T5, which have been used in anisotropy studies where it is advantageous to counterbalance the slightly reduced reconstruction accuracy with larger statistics~\cite{Aab:2017tyv,Aab:2018mmi,Aab:2020xgf}.
The whole acquisition chain, starting from the single-station trigger up to the off-line event selection, reduces the counting rate of single stations from about $3000/\text{s}$, due mainly to single uncorrelated atmospheric muons~\cite{Abreu:2011zza}, down to about $3{\times}10^{-5}/\text{s}$ for high quality physics events.
Consequently, a rate of events across the complete array of about $0.05/\text{s}$ is measured, which is almost entirely due to extensive air showers.

The footprint of one such event is shown in \cref{f:event} and is used in the examples that follow in later sections.

\section{Reconstruction at the level of the station}
\label{s:stationrec}

For recorded events, the treatment of signals starts at the station level.
In this section, we describe the treatment of the FADC traces\footnote{~A few traces (${<}3\%$) can in fact show anomalies due to hardware problems either in the PMTs or in the electronics: these are identified prior to the analysis by means of the monitoring data and excluded from the reconstruction.} for each PMT in stations passing the T1 or T2 trigger conditions.
These procedures provide a common standard to which all surface detector stations are calibrated.
Additionally, they provide a start time and a signal magnitude.
The uncertainties related to each of these are also discussed.

\subsection{Calibration}
\label{s:calibration}

\begin{figure}[t]
\centering
\includegraphics[width=0.5\columnwidth]{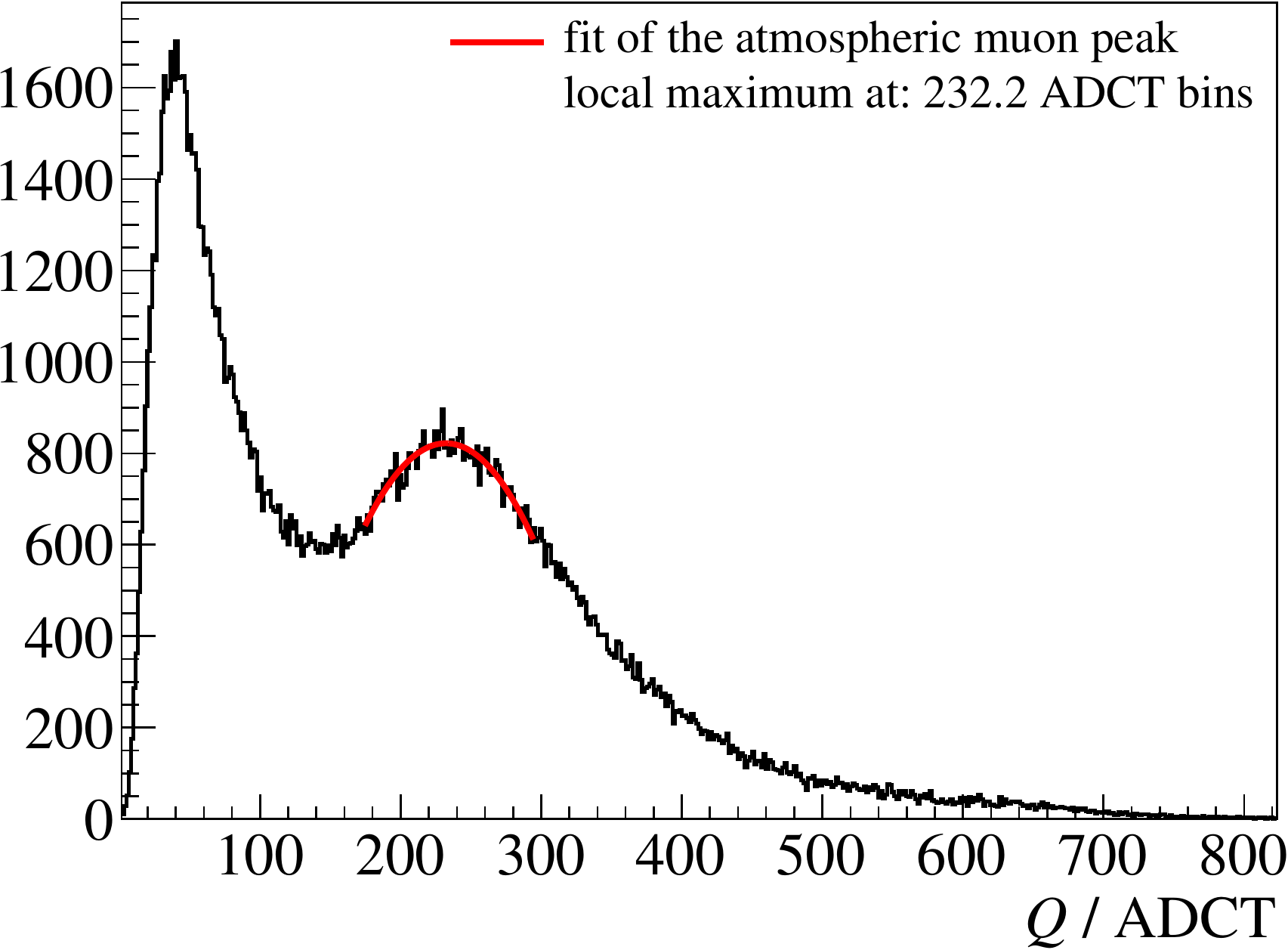}
\caption{The calibration of individual stations is performed by collecting 60\,s of background signals.
The histogram above is an example of the sizes of integrated signals $Q$ in units of ADCT = ADC count $\times$ time bin.
The position of the peak (local maximum) at $Q={\sim}232$\,ADCT, created by the atmospheric background muons, is used as the reference value for conversion of station signals into units of VEM.}
\label{f:charge_histo}
\end{figure}

The aim of the calibration is to enable a conversion of the signals recorded by the three PMTs and electronics of each station from the unit of ADC-counts to the reference unit of a vertical equivalent muon (VEM). The VEM is the charge  associated with a vertical muon passing through the center of the station and also corresponds to ${\sim}240$\,MeV of energy deposited due to ionization losses of the passing muon.~\cite{Bertou:2005ze}.
This conversion factor, $Q^\text{peak}_\text{VEM}$, is evaluated online by constantly measuring the properties of particles entering the detector volume at a rate of ${\sim}3/\text{s}$.
The on-board SD electronics are used to record histograms of the digitized pulse-heights and integrated charges from a minute's worth of background particles.
An example of such a charge histogram is shown in \cref{f:charge_histo}.
These histograms are stored alongside and exported with event data, thereby enabling the off-line re-evaluation of the VEM unit with 3\% accuracy.

The calibration factor $Q^\text{peak}_\text{VEM}$ is the position of the local maximum corresponding to the most probable charge deposited by muons coming from all directions, which is proportional to the VEM.

In each event, the FADC traces are independently calibrated for each PMT using the background data recorded in the preceding minute.
Dynamic station-by-station conversion of all signals into units of VEM facilitates preservation of a stable and uniform response to shower particles across the array.

\begin{figure}[t]
\centering
\includegraphics[width=\textwidth]{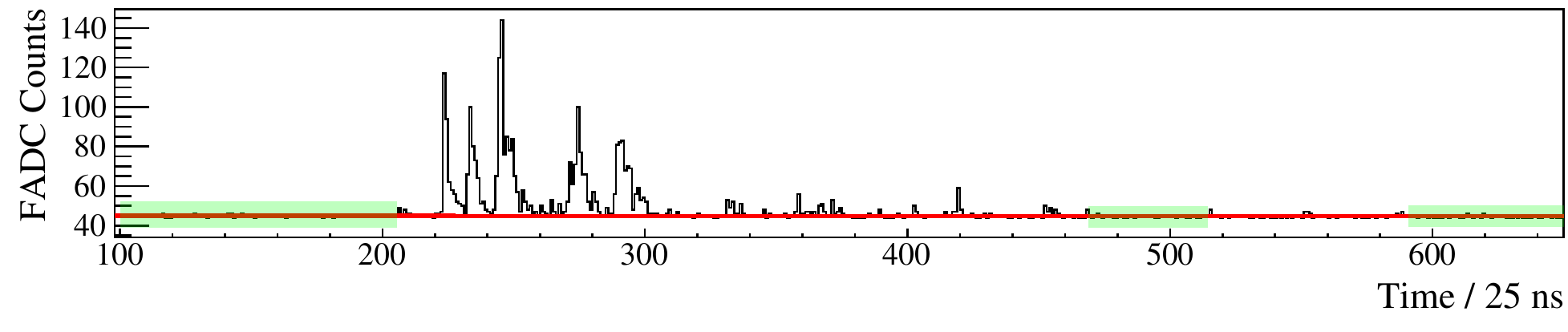}\\
\includegraphics[width=\textwidth]{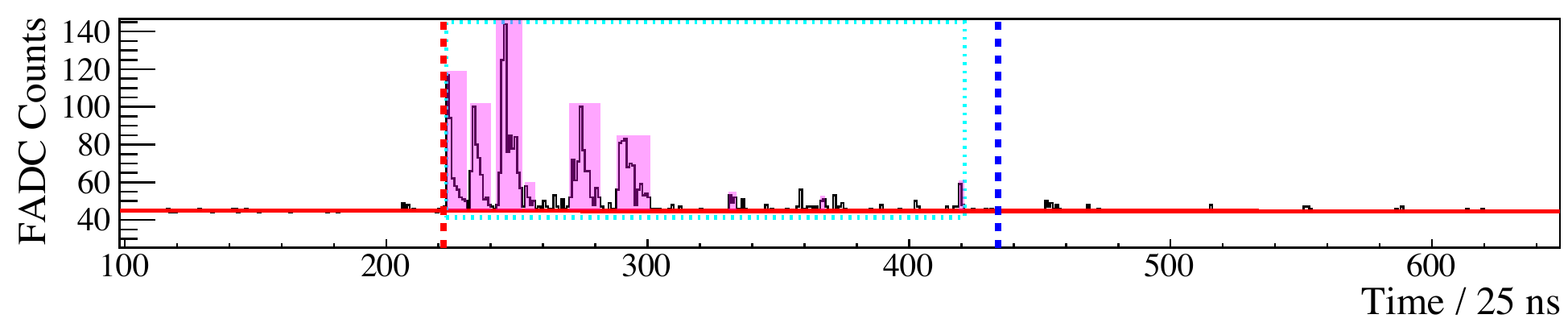}
\caption{\emph{Top:} Illustration of the baseline estimation (red), constructed from the individual segments (highlighted in green) and interpolation, on a high-gain FADC trace of a PMT.
\emph{Bottom:} Identification of signal fragments (magenta filled boxes) and the final merged signal window (large dotted cyan box).
Based on the number of peaks, this trace probably contains ${\sim}5$ muons with some additional electromagnetic component.
Trace start and stop times are denoted with the vertical dashed red and blue lines, respectively.}
\label{f:BaselinePieces}
\end{figure}

\subsection{Determination of the baseline}
\label{s:baseline}

The first step in the signal processing is the estimation of the baseline of the high-gain and low-gain traces of each PMT separately.
Since the baseline of a trace can be affected by the signals themselves, it is not estimated as one single value for the whole trace. Instead, the algorithm allows for a changing baseline along the trace.
Constant baseline segments are identified as at least 40 consecutive bins (i.e.\ with a duration of at least 1\,$\upmu$s) that vary in amplitude by less than $\bar{y}\pm\sigma_\text{b}$, where $\bar{y}$ is the mean of the baseline segment and $\sigma_\text{b}$ is the tolerance needed to account for the possible noise in the baseline. $\sigma_\text{b}$ is set to 2\,ADC counts.
In rare cases (${<}0.2$\%), when the PMT in question is noisy, the tolerance $\sigma_\text{b}$ is increased in steps of 1 until a constant baseline segment is found.
These baseline segments, illustrated by the green boxes in \cref{f:BaselinePieces}-top are then combined to form the total baseline.
Each of the bins corresponding to a baseline segment is inserted into the total baseline as a constant value given by $\bar{y}$ of that baseline segment.
Then, the parts between the baseline segments are filled with values interpolated linearly in integrated signal, which results in an estimation of the baseline covering all of the 768 bins constituting the trace as shown with the red line in \cref{f:BaselinePieces}.

\subsection{Identification of the time range of interest}
\label{s:signal}

The next step is the extraction of the relevant signal from the traces of individual PMTs, i.e.\ identifying the start and stop time of the signal, which is always performed on the high-gain channel due to its superior resolution.
As in the baseline case, the FADC traces are first scanned to identify candidate signal \emph{fragments}, which consist of consecutive bins with amplitudes of at least 3\,ADC counts above the baseline.
Examples of identified signal fragments are indicated in \cref{f:BaselinePieces}-bottom with magenta boxes.
After the subtraction of the baseline in each PMT, traces are calibrated in VEM units using the value of $Q^\text{peak}_\text{VEM}$.

To avoid inclusion of background signals (mostly single, unaccompanied muons), the candidate fragments are additionally scrutinized before being merged into the final signal \emph{segment}.
Two consecutive fragments A and B can only be merged if they fulfill both of the following requirements:
\begin{enumerate}
\item Fragment B must begin less than 500\,ns (20 time bins) + $t_\text{A}$ after the end of fragment A, where $t_\text{A}$ is the amount of time in fragment A containing signal.
\item The integral of fragment A must be greater than 30\% of the integral of fragment B, or fragment B must have an amplitude below $5I^\text{peak}_\text{VEM}$, where $I^\text{peak}_\text{VEM}$ is the mean amplitude of the background muons~\cite{Bertou:2005ze}.
\end{enumerate}
This procedure is repeated for all consecutive fragments until all fragments are processed and signal segments are created.
In \cref{f:BaselinePieces}-bottom, the final signal segment resulting from the merging of the candidate fragments (magenta) is indicated with a dotted cyan box. 

The last step of the procedure consists in the averaging of the signal segments of the three PMTs (or of all working PMTs\footnote{~Currently 12\% (1\%) of stations have only two (one) functioning PMTs.}) into the station-level signal segment.

For the final station-level trace, only the segment with the largest signal is selected and the corresponding start and stop times are adopted with a minimal run of 250\,ns to ensure containment of any potential sub-threshold signals. 

With this start time, we obtain our best estimate of the beginning of the passing shower front.
With this stop time, we ensure that all particles belonging to the shower in question are included while excluding as many accidental signals as possible.

\subsection{Estimation of the signal}
\label{s:finalsignal}\label{s:saturation}

The total signal of a station is obtained by integrating the final trace, which consists of the bin-by-bin average of the high-gain (or low-gain, if the high gain is saturated) traces of the working PMTs between the previously determined start and stop times.

When the flux of particles entering a station is large (e.g.\ when a shower lands close to a station), the signal can saturate the ADC range of the high-gain channel.
This normally occurs when the equivalent integrated signal is greater than ${\sim}60$\,VEM.
When more than two FADC-trace bins reach the maximum ADC count ($2^{10}-1=1023$), the final station-level trace is constructed from the average of the low-gain PMT-traces instead, while the start and stop times of the high-gain signal are kept.

When the core lands even closer to a station, the low-gain channels can also saturate and the corresponding events are termed \emph{saturated events}.
For lower energy showers, this occurs in 10\% of events, for which  the distance of this station is closer than ${\sim}100$\,m to the shower axis.
For the highest energy showers, saturation already occurs when a station is closer than ${\sim}500$\,m, which occurs in more than 50\% of events (see \cref{s:s1000rec} for more details).
In the case of low-gain saturation, the equivalent integrated signal is typically greater than ${\sim}1000$\,VEM.
Events with more than one saturated station are extremely rare\footnote{~Out of the more than 400\,000 events, we have identified only three where two stations have saturated.}.

With the increasing signal, we first observe an overflow of the ADC range of the low-gain channels and then, with even larger signals, we observe saturation of the PMT itself (i.e.\ the departure from the linear behavior of the photomultiplication).
On such traces, a signal-recovery algorithm is applied~\cite{Veberic:ICRC13} to estimate the true magnitude of the low-gain signal from the amplitude of the signal undershoot (the dip of the trace below baseline).
In addition, in a more detailed study it has been shown that the non-linear response of the PMT, when measured precisely, can be used to recover the saturated signals by fitting the non-saturated parts of the trace.
For all PMTs for which the saturation and non-linearity response has been measured individually, the range of signal estimation can be increased up to $10^6$\,VEM, with a resolution of 20\%~\cite{Veberic:ICRC13}.
The latter recovery procedure is implemented only in the \emph{Observer} framework.

When the high-gain channel is used to produce the calibrated station trace, the high-gain signal is scaled with the value of $Q^\text{peak}_\text{VEM}$ since this is the channel with which muon data are continuously monitored by the online station calibration.
In cases where the high-gain channel saturates, the low-gain channel is used to create the calibrated station trace.
In this case, the low-gain signal is scaled with the product of the $Q^\text{peak}_\text{VEM}$ factor and the high-to-low-gain ratio, which is also constantly monitored by the online station calibration.

\subsection{Uncertainties in the measurement of the signal}
\label{s:signaluncertainty}

\begin{figure}[t]
\def\figh{0.40}
\centering
\includegraphics[height=\figh\columnwidth]{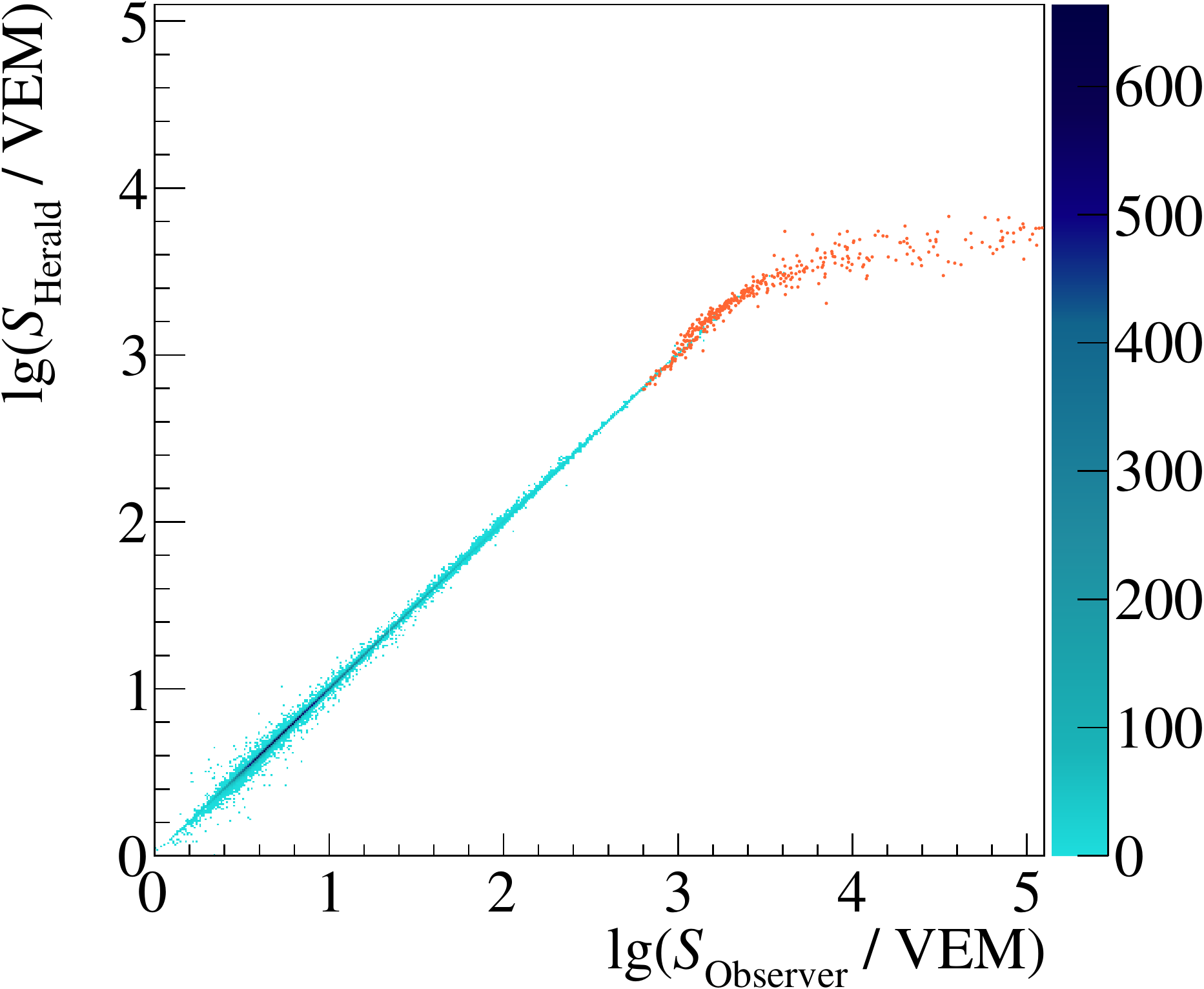}\hfill
\includegraphics[height=\figh\columnwidth]{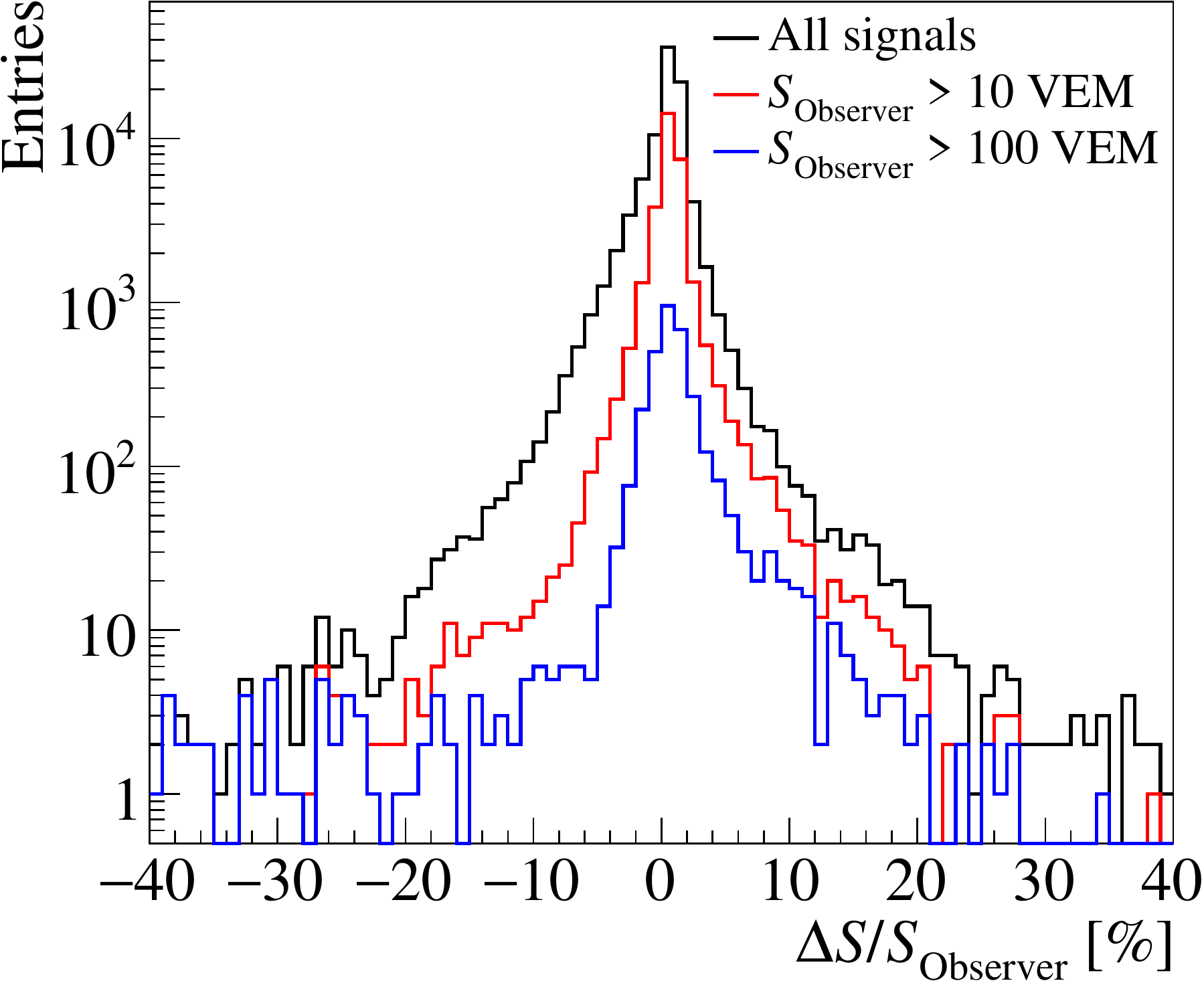}
\caption{Comparison of the signal distributions of the two reconstructions. 
\emph{Left:} Below 1000\,VEM, above which the low-gain signal of some stations begins to saturate, the distribution can be fitted with the following power law $S_\text{Herald}/\text{VEM}=1.02(S_\text{Observer}/\text{VEM})^{0.98}$.
The saturated stations are represented by the orange points.
\emph{Right:} Distribution of relative signal difference $\Delta S/S_{\text{Observer}}$, where $\Delta S=S_\text{Herald}-S_\text{Observer}$, for different ranges of signal.
For all ranges of signals, the average of the distribution is below 1\%, and the standard deviation ranges from less than 2\% to 5\% for the full data set.}
\label{f:difsig}
\end{figure}

The uncertainties on the integrated signal and on the start time of the signal are obtained from studies using configurations with \emph{twin} stations, i.e.\ pairs of stations placed only ${\sim}11$\,m apart and thus sampling the same part of a shower.
In the studies laid out below we used all available data recorded by twins.
Two twins were already deployed in 2003 and remain operational today.
In 2006, 20 additional twins were added.
Of these, 13 collected data for nearly 5 years, while a remaining hexagon of seven twins remains operational.

The uncertainty in the start time strongly influences the angular resolution of the reconstruction of the shower geometry.
This uncertainty is composed of the time resolution of the GPS time-tagging system (${\sim}10$\,ns), the 40\,MHz FADC sampling (${\sim}7$\,ns), and, more importantly, by the fluctuations of the arrival times of the first particles.
These depend on the number of particles entering the station, i.e.\ a convolution between the thickness of the shower front, the density of particles therein, and the cross-section of the detector. 
At large radial distances only a few particles will enter a detector and thus large start-time fluctuations can occur due to this sparse sampling of the shower front.
We can describe the effects above with a general time variance model~\cite{Bonifazi:2007ck} for which the uncertainty of the start time $t_\text{start}$ is estimated as
\begin{equation}
\sigma_{t_\text{start}}^2 =
a^2\left(\frac{2 t_{50}}{n}\right)^2\frac{n - 1}{n + 1} + b^2 ,
\label{eq:time_variance}
\end{equation}
where $t_{50}$ is the length of the time interval that contains the first 50\% of the total signal, and the effective number of particles $n$ is obtained from the signal $S$ as $n=S/\bar{\ell}(\theta)$, where $\bar{\ell}(\theta)$ is the mean relative track-length of through-going particles crossing the detector when arriving with a zenith angle $\theta$.
Due to the cylindrical shape of the water-Cherenkov tank, $1/\bar{\ell}(\theta) = \cos\theta + (2H/\pi R)\sin\theta$, where $H$ and $R$ are the height and the radius of the cylinder, respectively.
The coefficients $a$ and $b$ (with typical magnitudes of 0.7 and 16\,ns, respectively) are parameterized as functions of the zenith angle of the shower and were determined using twin stations.

For the signal uncertainty of non-saturated stations, a Poisson-like parameterization,
\begin{equation}
\sigma^2_S = f^2_S(\theta)\,S, \quad \text{where} \quad f_S(\theta) = 0.34 + 0.46\,\sec\theta,
\label{eq:sigma_S}
\end{equation}
is a good description of the observed signal fluctuations.
The values from~\cite{Ave:2007zz} have been updated as a result of studies with larger statistics.

However, for stations with saturated PMTs, the accuracy of the signal measurement is degraded. 
In the \emph{Observer} framework, the signal uncertainties are increased by the intrinsic uncertainties of the saturation-recovery algorithm.
The signal uncertainty for the recovered signals varies from 5\% at ${\sim}1000$\,VEM up to 60\% at ${\sim}10\,000$\,VEM.
For very large signals, i.e. above 10\,000 VEM, the PMT response is highly non-linear, and only few measurements of the saturation curve at large anode currents are available.
The algorithm uses average curves, which results in uncertainties greater than 50\% on the recovered signal.
In the \emph{Herald} framework, the algorithm for signal recovery does not take into account the PMT non-linearity and therefore only the timing information is used for signals larger than 2000\,VEM.

For both the determination of the timing and the estimation of the signal and its uncertainties, the two reconstruction frameworks exhibit some differences.
Nevertheless, this has only a limited impact on the final estimate of the signal size.
The differences between the signals are on average less than 1\% with a standard deviation varying from 2\% to 4.9\%.
As can be seen in \cref{f:difsig}-right, they are mostly observed at the lowest signals, and are within the estimated uncertainties.
For the saturated signals, the large differences observed in \cref{f:difsig}-left are explained by the additional recovery procedure used in the \emph{Observer} framework.
The impact of saturation on the event reconstruction is discussed later in this paper.

\section{Building events}
\label{s:evebuilding}

Prior to reconstruction, each event is built from the data gathered by the T3 trigger.
The sizes and start times of the signals of the triggered stations are inspected to identify candidate showers.
Details of the selection process can be found in~\cite{Abraham:2010zz}.
Here we summarize only the elements most relevant for the reconstruction.

Since the highly sensitive SD stations may be triggered due to lightning during thunderstorms, affected events must be identified and removed.
This is done using a Fourier-like algorithm to detect oscillations in traces, the presence of which indicates that lightning has struck nearby.
If even a single PMT of any station participating in the event is found in this condition, the whole event is discarded.

Next, a selection process is applied to the triggered stations of candidate shower events to discard accidental stations, i.e.\ stations which triggered by chance and are not, in fact, part of the event.
To perform such a selection, an estimate of the time the shower front was expected to have passed through each station is required.
This is obtained by means of a first order \textit{seed} reconstruction, which, in addition to facilitating the identification and exclusion of accidental stations, also provides the initial estimates for the geometric parameters of the proper event reconstruction detailed in \cref{s:everec}.

\begin{figure}[t]
\centering
\includegraphics[width=\textwidth]{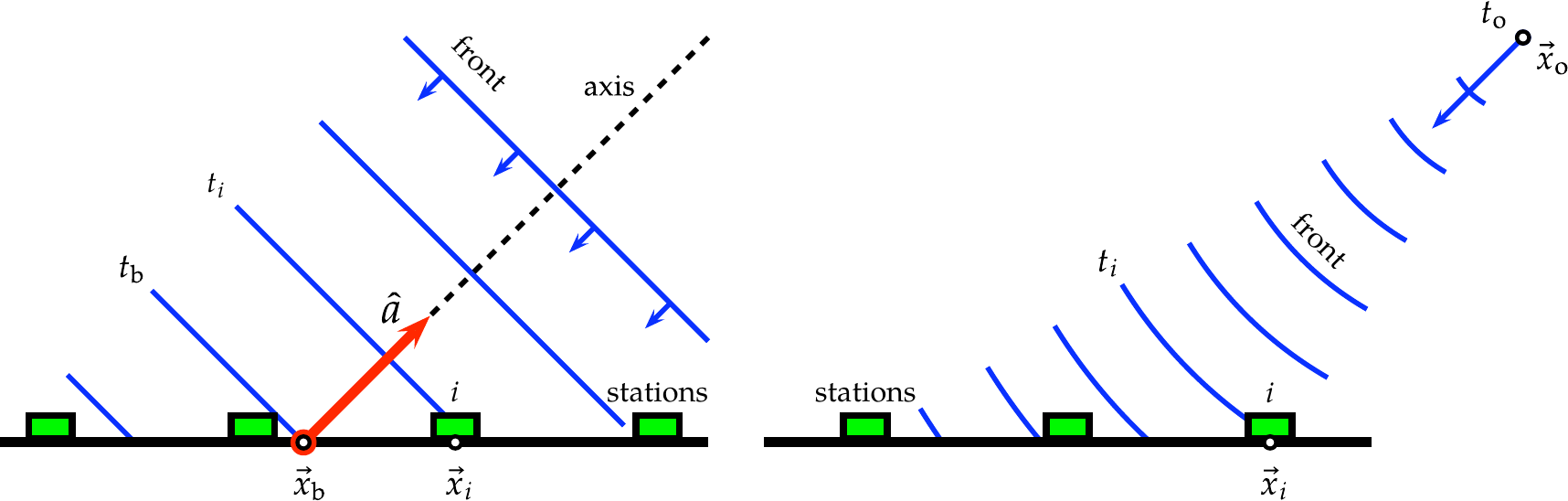}
\caption{\emph{Left:} Schematic description of the plane-front approximation for air-shower development (see text).
\emph{Right:} Schematic of the spherical shower-front development.
The apparent origin of the shower is at point $\vec{x}_\text{o}$ at time $t_\text{o}$.
The shower front reaches the station $i$ at position $\vec{x}_i$ at time $t_i$.}
\label{f:plane}
\label{f:sphere}
\end{figure}

The seed reconstruction is obtained as follows.
The to-be-determined axis $\hat{a}$ of the shower is anchored at the location of a barycenter $\vec{x}_\text{b}$ (i.e.\ the signal-weighted\footnote{~Monte Carlo studies have shown that weights proportional to the square-root of the signal in the stations result in a barycenter on average closest to the impact point of the shower core on the ground.} center-of-mass of stations in an event), which later\footnote{~Note that at this stage the axis $\hat{a}$ can be obtained without any reference to the shower core.} serves as a first estimate of the impact position of the shower core at the ground.
The assumed shower core travels in the $-\hat{a}$ direction, impacting the ground at barycenter $\vec{x}_\text{b}$ at the time $t_\text{b}$.
Under the assumption that particles in the shower front move in a plane perpendicular to the shower axis with the same speed as the core of the shower (see \cref{f:plane}-left), which is assumed to be the speed of light $c$, the time $t_\text{sh}(\vec{x})$ when the shower plane passes through some arbitrary point $\vec{x}$ (e.g.\ a station on the ground) may be inferred through a simple projection onto the shower axis as
\begin{equation}
c\,t_\text{sh}(\vec{x}) = c\,t_\text{b}-\hat{a}\cdot(\vec{x}-\vec{x}_\text{b}).
\label{shower-timing}
\end{equation}
The solution to this equation is obtained by identifying a reconstruction \emph{seed triangle} consisting of three stations: a station with its two nearest-neighbors in a non-aligned configuration, where all three stations must have passed a station level trigger.
The seed triangle with the highest sum of the three station signals is used to analytically determine an exact solution, thereby obtaining an estimate of the shower axis and an estimation of when the shower front arrived at any position under the plane-front assumption.

We designate as accidental all stations where the onset of the signal is more than $1\,\upmu$s prior to (early) or $2\,\upmu$s after the expected passage of the plane shower front (late), which ensures that more than 99\% of the stations belonging to the shower are kept in the event.
The criteria is more strict for early stations since the seed triangle, which defines the shower front estimate, is based on stations close to the shower axis with the highest signals, which are more likely to attain signal already at the start of the passage of the shower front.
Stations more distant from the shower axis measure smaller signals whose start times are biased to later times due to the larger sampling fluctuations in the arrival time of the first particles as well as due to the curvature of the shower front resulting in increasing delay with respect to the plane shower front.
Additionally, stations are removed from an event when there are no triggered stations within 1.8\,km (or 5\,km) for a single station (or two neighboring stations).
Furthermore, active stations which did not record enough signal to trigger and are located up to 5\,km from the shower core are considered as part of the event.
They provide additional constraints on the location of the impact point, as will be discussed in the next section.

\section{Reconstruction of events}
\label{s:everec}

The timing and the size of the signal measured in each station, as well as accurate knowledge of the 3D positions of the stations, are the key inputs for the reconstruction of the arrival directions and the sizes of the showers selected according to the criteria described in the \cref{s:evebuilding}.
To reconstruct these quantities, we adopt a simplified model of air showers that allows us to separate the process into two parts.
First, from the timing information of the stations, we determine the geometry of the shower, namely the direction of the shower axis $\hat{a}$ and the position of the impact point of the shower core on the ground $\vec{x}_\text{c}$.
Using this geometry, the second step consists of fitting the signal magnitudes as a function of distances from the shower axis to an empirically-derived functional form describing the average lateral distribution of particles.
These two steps can be formally intertwined and thus require iteration or a global fit, but can also be effectively separated with a particular choice of the shower front.

\subsection{Shower geometry}
\label{s:geom}

From the selection of the stations described in \cref{s:evebuilding}, the three stations of the seed triangle are used to give the first rough estimate of the axis $\hat{a}$ of the shower and the impact position $\vec{x}_\text{c}$ of the core on the ground.
The shower models used here describe the secondary particles that ``have traveled furthest in the forward direction'', also referred to as the \emph{extreme front}~\cite{Linsley:1962kq}, as moving with the speed of light in a curved shower front.
The radius of curvature is added as a free parameter when five or more stations are participating in the event.
Moreover, for the reconstruction of the shower geometry, we consider the core to be moving in the $-\hat{a}$ direction and intersecting with the ground at the impact point of the shower core $\vec{x}_\text{c}$ at the time $t_\text{c}$.
Thus four parameters are fitted: the two directional cosines\footnote{~Here we try to avoid fitting the angles $\theta$ and $\phi$ of the spherical coordinate system due to the singularity at the zenith $\theta=0$. Nevertheless, they are used in the fits of more inclined events with zenith angles $\theta>60^\circ$, c.f.~\cite{Aab:2014gua}.} $u$ and $v$ of the unit vector $\hat{a}=(u,v,\sqrt{1-u^2-v^2})$, the time $t_\text{c}$, and the radius of curvature of the shower front $R_\text{o}$.
Using this model for the arrival time of the shower front, $t_\text{sh}(\vec{x})$, the shower geometry is fitted to the start-time of the signals $t_i$ in each triggered station $i$ located at $\vec{x}_i$ using the \textsc{Minuit} framework~\cite{minuit}.
The function which is minimized is the sum of the squares of the differences in predicted and measured start times,
\begin{equation}
\chi^2 = \sum_i\frac{[t_i-t_\text{sh}(\vec{x}_i)]^2}{\sigma^2_{t_i}},
\label{eq:chi2timing}
\end{equation}
where $\sigma_{t_i}$ is the uncertainty of the start time $t_i$, as given by \cref{eq:time_variance}, which derives, in part, from fluctuations in the arrival times of the most forward particles in the shower front at a station $i$.
The absolute positions of the stations $\vec{x}_i$ were measured with the built-in GPS receivers in dedicated campaigns to a precision of around 20\,cm horizontally and 50\,cm vertically so that the positional uncertainty has only a negligible effect on the event reconstruction.

As initial values for the fitting procedure, the signal-weighted barycenter $\vec{x}_\text{b}$ from \cref{s:evebuilding} is chosen as a suitable approximation for the impact point $\vec{x}_\text{c}$, while the result of the plane fit from \cref{shower-timing} is used for the shower axis $\hat{a}$.

The two reconstruction frameworks include slightly different descriptions of the shower front curvature.
In the \emph{Herald} framework, the curvature is considered constant, with the particles of the shower are distributed on a spherical front moving along the direction of $-\hat{a}$.
Compared to the plane front in \cref{shower-timing}, the particles are thus delayed proportionally to  $R_\text{o}-\sqrt{R_\text{o}^2-[r_{\hat{a}}(\vec{x}-\vec{x}_\text{c})]^2}$, where $R_\text{o}$ is the constant radius of the spherical front and where
\begin{equation}
r_{\hat{a}}(\vec{x}) = |\hat{a} \times \vec{x}|
\label{eq:r}
\end{equation}
is the perpendicular distance of point $\vec{x}$ from the shower axis $\hat{a}$.
To keep $t_\text{sh}(\vec{x})$ linear in the fitted parameters, only terms up to the second order in the $r/R_\text{o}$ expansion are used.
With a curvature parameter $k_\text{o}=1/2R_\text{o}$, we can then express the shower timing as
\begin{equation}
c\, t_\text{sh}(\vec{x}) =
    c\,t_\text{c}
  - \vec{a}\cdot(\vec{x}-\vec{x}_\text{c})
  + k_\text{o} \, [r_{\hat{a}}(\vec{x}-\vec{x}_\text{c})]^2,
\label{eq:timeCDAS}
\end{equation}
which is clearly a \emph{paraboloidal} extension\footnote{~The size of the quadratic term is typically on the order of ${\sim}500$\,ns at distances of ${\sim}2500$\,m.} of \cref{shower-timing}.

The \emph{Observer} reconstruction approximates the shower development as starting at time $t_\text{o}$ from a virtual point of origin $\vec{x}_\text{o}$ (see \cref{f:sphere}-right) and propagating towards the ground in the shape of a \emph{sphere}, concentrically inflating with the speed of light\footnote{~The typical difference in time between the paraboloidal and the spherical description of the shower plane is of the order of 50\,ns at distances of ${\sim}1500$\,m.}.
The arrival time of such a shower front at a point $\vec{x}$ is thus
\begin{equation}
c\,t_\text{sh}(\vec{x}) = c\,t_\text{o} + |\vec{x}-\vec{x}_\text{o}|.
\label{eq:timeOffline}
\end{equation}
Note that contrary to the paraboloidal description in \cref{eq:timeCDAS}, this spherical fit can be performed without any prior knowledge of the impact point $\vec{x}_\text{c}$ or the shower axis.
Once $\vec{x}_\text{c}$ is determined at a later point, the shower axis may be obtained as a normalized direction towards $\vec{x}_\text{o}$ as $\hat{a}=(\vec{x}_\text{o}-\vec{x}_\text{c})/|\vec{x}_\text{o}-\vec{x}_\text{c}|$.
Due to the expansion of the sphere used in this model, the radius of curvature of the shower front depends on time, $R(t)=c(t-t_\text{o})$.
Nevertheless, for consistency with \cref{eq:timeCDAS}, the radius of curvature is defined as the distance between the virtual origin and the impact point, $R_\text{o}=|\vec{x}_\text{o}-\vec{x}_\text{c}|$, i.e.\ as the radius of the shower at the time of passage $t_\text{c}=t_\text{o}+R_\text{o}/c$ through the impact point $\vec{x}_\text{c}$.

For both of the models above, there are four free parameters to describe the development of the shower front.
Since low-energy events have a station multiplicity of only three or four, they do not have enough degrees of freedom in the timing data to solve for the shower-front curvature.
For events with less than five triggered stations, we therefore use the curved model with an $R_\text{o}$ fixed to a parametrization optimized using events with a larger number of stations.

\begin{figure}[t]
\def\figh{0.34}
\centering
\includegraphics[height=\figh\columnwidth]{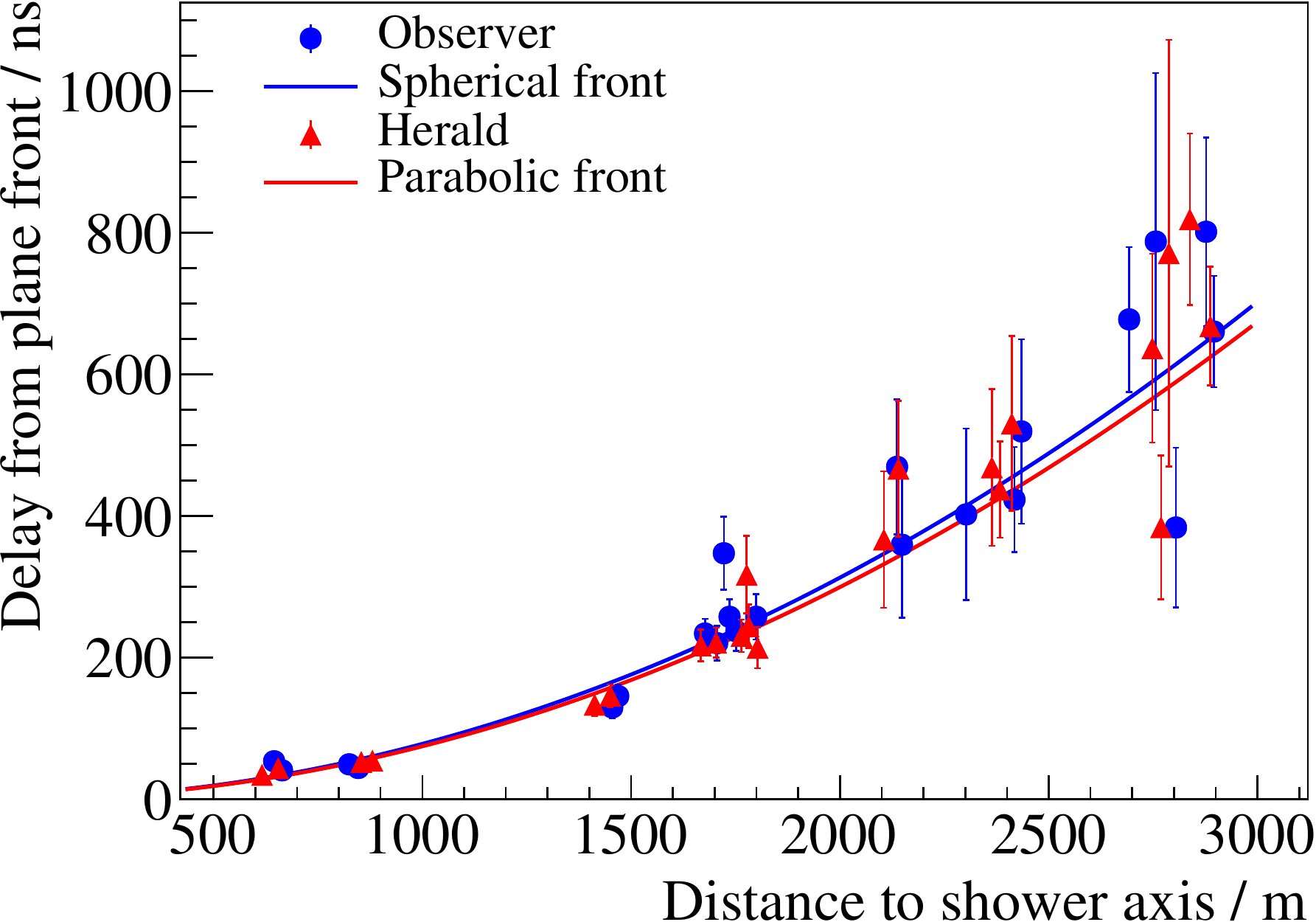}\hfill
\includegraphics[height=\figh\columnwidth]{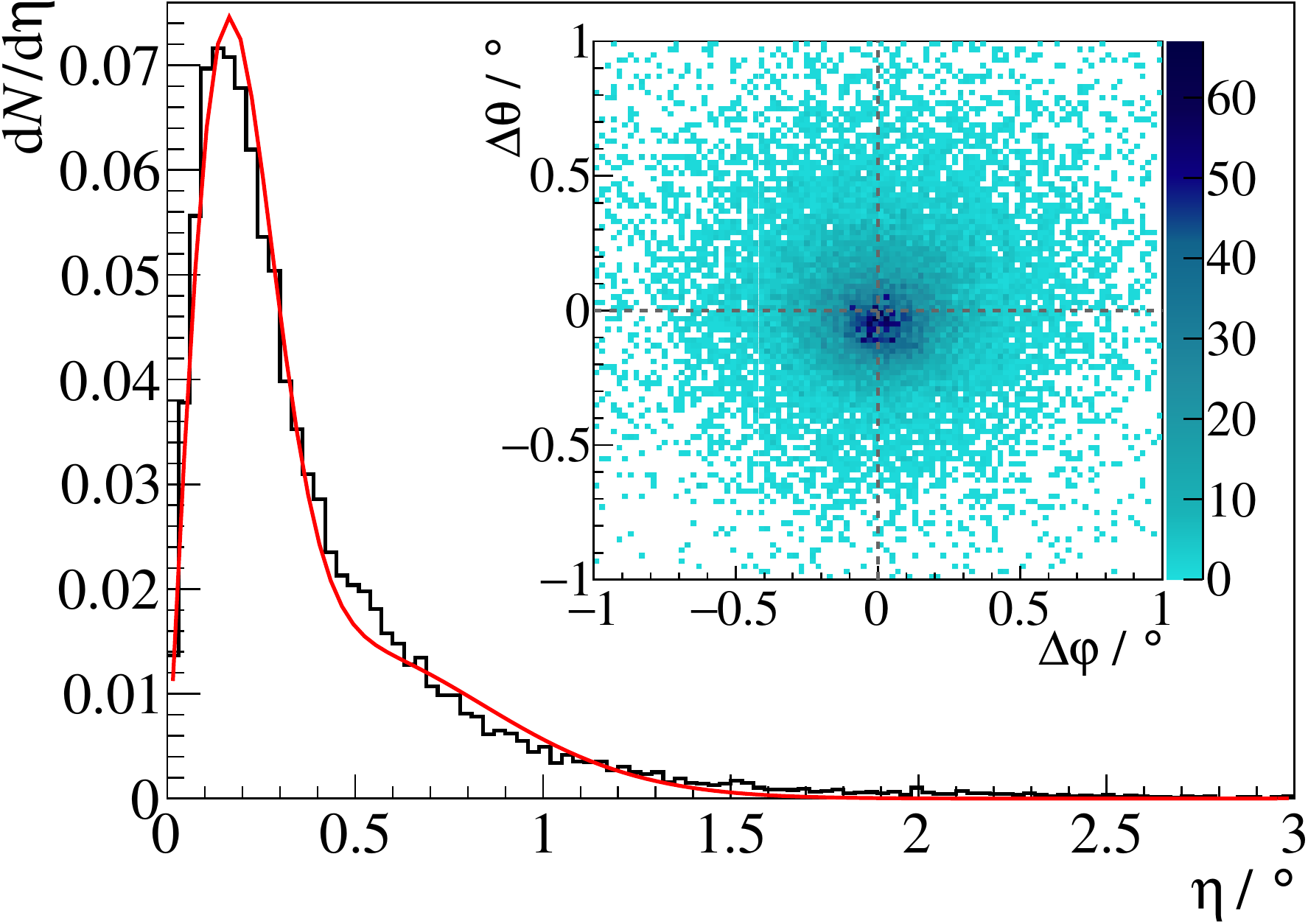}
\caption{\emph{Left:} For the large example event from \cref{f:event}, the start times in individual stations are shown as a function of the distance to the shower axis for the \emph{Herald} (red triangles) and \emph{Observer} (blue circles) reconstructions.
The times are relative to the arrival time of the corresponding plane front.
The solid lines represent the predicted time for the spherical model (\cref{eq:timeOffline}, blue) and the parabolic model (\cref{eq:timeCDAS}, red).
\emph{Right:} Distribution of the angular difference $\eta$, where $\sin\eta=|\hat{a}_\text{p}\times\hat{a}_\text{s}|$, between the axes of the two curved shower-front models: $\hat{a}_\text{p}$ for the paraboloid and $\hat{a}_\text{s}$ for the spherical model.
The distribution has been fitted with a weighted sum of two Rayleigh distributions from which we can derive that 68\% of reconstructed events have arrival directions which agree to within $0.40^\circ$.
\emph{Inset:} 2D distribution of the differences in the zenith angle, $\Delta\theta=\theta_\text{s}-\theta_\text{p}$ and in the azimuthal angle, $\Delta\varphi=\varphi_\text{s}-\varphi_\text{p}$.
}
\label{f:timing_angle}
\end{figure}

As an example, a geometrical fit of the example event from \cref{f:event} is shown in \cref{f:timing_angle}-left for both reconstructions.
The existence of curvature in the shower front can be clearly observed in the increasing delays of signal start-times with respect to the arrival of a plane front tangential at the shower core. 

The angular differences between the axes of the two reconstructions are shown in \cref{f:timing_angle}-right, from which we can conclude that 68\%, 90\%, and 95\% of all the events are reconstructed within an angular difference of less than $0.40^\circ$, $0.83^\circ$, and $1.14^\circ$, respectively.
The small, non-zero difference of less than $0.1^\circ$ observed in $\Delta\theta$ will be addressed in \cref{s:ldf}.

\subsection{Shower size}
\label{s:shower_size}

The shower size is estimated with a regression of the parameters of the lateral distribution function (LDF), $S(r)$, to the signals in the triggered stations and is additionally constrained by the absence of triggers in stations which would have measured very little or no signal.
The LDF is a function of the perpendicular distance to the shower axis, $r$.
However, showers induced by identical primaries (i.e.\ with the same energy, mass, and arrival direction) can be sampled at the ground at different stages of development due to the shower-to-shower fluctuations arising from the variability of the location and the nature of the leading interactions, and the variability of the subsequent development of the cascade.
This results in a natural variability of the shower size, the mean value of which depends on the mass of the primary particle.

For the majority of events, the multiplicity and spatial distribution of triggered stations in the shower plane is insufficient to precisely estimate the shower size and determine the shape of the LDF.
Instead, station signals of individual events are fitted with a scaled, data-derived \emph{average} LDF shape $f_\text{LDF}(r)$ such that
\begin{equation}
S(r) = S(r_\text{opt}) \, f_\text{LDF}(r),
\label{ldf}
\end{equation}
where $S(r_\text{opt})$ is the shower-size estimator and $f_\text{LDF}(r)$ is normalized so that $f_\text{LDF}(r_\text{opt})\equiv1$.
The optimal value for distance $r_\text{opt}$ has been chosen so that the variability of the shower-size estimator $S(r_\text{opt})$ with respect to the aforementioned shower-to-shower fluctuations is minimized.
The value almost exclusively\footnote{~$r_\text{opt}$ depends only at ${\sim}5\%$ level on the choice of the LDF form, the primary energy, and the zenith angle.} depends on the spacing and structure of the array~\cite{Newton:2006wy}.
In the case of the Auger SD, where the array is an isometric triangular grid with a spacing of $1500$\,m, the optimal distance amounts to $r_\text{opt}\approx1000$\,m, so we have decided to fix it at exactly 1000\,m and denote the corresponding shorthand for the shower-size estimator as $S(1000)$.

\subsubsection{Lateral distribution function}
\label{s:ldf}

Due to the lack of analytical solutions for the hadronic-cascade equations, functional forms for $f_\text{LDF}(r)$ have traditionally been chosen empirically (see~\cite{Nagano:2000ve} for an overview of functional forms chosen for previous experiments).
The \emph{Herald} reconstruction uses a log-log parabola,
\begin{equation}
\ln f_\text{LDF}(r) = \beta\,\rho + \gamma\,\rho^2,
\label{eq:loglog}
\end{equation}
where $\rho=\ln(r/r_\text{opt})$ and $\beta$, $\gamma$ are the average slope parameters of the LDF.
For smaller axial distances, $r<r_\text{c}$ where $r_\text{c}=300$\,m, a tangential log-log linear function $\beta\,\rho+\gamma(2\rho-\rho_\text{c})\rho_\text{c}$ is used where $\rho_\text{c}=\ln(r_\text{c}/r_\text{opt})$.

The \emph{Observer} reconstruction instead uses a slightly modified NKG function~\cite{Kamata58,Greisen56,Greisen:1960wc} of the form
\begin{equation}
f_\text{LDF}(r) =
  \left(\frac{r}{r_\text{opt}}\right)^\beta
  \left(\frac{r+r_\text{s}}{r_\text{opt}+r_\text{s}}\right)^{\beta+\gamma},
  \label{eq:nkg}
\end{equation}
with a fixed $r_\text{s}=700$\,m.

The average of the slope parameter $\beta$ is determined in both cases in a data-driven way, by fitting $\beta$ for the subset of events with a multiplicity and spatial distribution of stations providing a sufficient \emph{lever arm}.
In these events, the slope $\beta$ is fitted given that there are at least 2, 3, or more stations within a radius of $400<r/\text{m}<1600$ and that at least two of the stations are separated by 900, 800, or 700\,m, respectively.
This lever-arm criteria ensures that the lateral distribution is sufficiently sampled around the distance of 1000\,m to constrain the slope of the LDF well.

Due to the strong correlation between $\beta$ and $\gamma$, it is more difficult to determine the average of the slope parameter $\gamma$ in a data-driven way in the case of both LDF functions.
In \cref{eq:loglog,eq:nkg}, the parameter $\gamma$ is describing the deviation of the LDF from a simple power-law function at large distances.
In the \emph{Observer} reconstruction, similar lever-arm criteria are used to fit $\gamma$ except the station search radius is changed to $1000<r/\text{m}<2000$.
In the \emph{Herald} reconstruction, the parameter $\gamma$ has been determined from Monte-Carlo studies.
Finally, both parameters $\beta$ and $\gamma$ are parametrized as functions of the zenith angle and $S(1000)$.

Both LDF models used within the Auger Collaboration assume that the deposited signals in the stations are rotationally symmetric around the shower axis.
In truth, the showers are asymmetric due to a combination of the longitudinal evolution and geometrical effects related to the angles of incidence of the particles on the stations.
The \emph{Herald} reconstruction corrects for these asymmetries with an unbiasing model applied to the measured signals
\begin{equation}
S_\text{symm}(r) = \frac{S(r)}{1 + \alpha(r)\cos\zeta},
\label{eq:azimuthalAsymm}
\end{equation}
where $\zeta$ is the azimuthal angle of the station (as measured in the shower plane where $\zeta=0$ corresponds to the upstream direction) and where the radial dependence of the amplitude $\alpha(r)$ has been obtained from simulations.
These corrections are only important in very specialized studies as they only result in small shifts of ${\sim}40$\,m in the estimate of the position of the shower core, which is well within the typical uncertainties of the reconstruction.
These corrections ultimately have a ${<}0.2\%$ effect on the value of $S(1000)$ and a ${<}0.2^\circ$ effect on the axial direction, as estimated by performing the \emph{Herald} reconstruction with and without their application.
The exact magnitudes of these effects depend on the zenith angle and state of development of the shower.
Since no asymmetry correction is applied in \emph{Observer}, the corresponding shift of the impact-point is also responsible for the very small, yet systematic differences in arrival directions reconstructed by the two reconstructions, where $\Delta\theta<0.1^\circ$ as seen in \cref{f:timing_angle}-right.

\subsubsection{Maximum-likelihood procedure}
\label{s:s1000rec}

Using the results of the reconstruction of the shower geometry, the fit to the LDF is, in the next step, maximized using the likelihood $\mathcal{L}$, composed as the product of probabilities $P$ over the coordinates of the shower impact point $\vec{x}_\text{c}$ and the size $S(1000)$, given the observed signal sizes $S_i$ in the stations, located at $\vec{x}_i$.
We are thus maximizing the log-likelihood
\begin{equation}
\ln\mathcal{L} =
  \sum_i\ln P(S(1000),\vec{x}_c\,|\,S_i,\vec{x}_i).
\label{basicLikelihood}
\end{equation}

The two reconstructions use different models for calculation of the likelihood $P$.
In the \emph{Observer} framework, $P$ is a product of contributions from: (a) stations with \emph{small} signals, (b) stations with \emph{large} signals, (c) stations with \emph{saturated} signals, and (d) \emph{non-triggered} stations.
The individual terms are as follows:
\begin{itemize}

\item[(a)] \emph{Small signals}.
For a small signal $S_i<S_\text{stat}$ where $S_\text{stat}\approx20$\,VEM, the likelihood is modeled as a Poissonian probability $f_\text{Poi}(S_i,S(r_i))$, with the signal scaled to a particle count for which the corresponding Poisson variance matches the signal variance from \cref{eq:sigma_S}, and where $S(r_i)$ is, according to the LDF of \cref{ldf}, the expected signal at the radial distance $r_i=r_{\hat{a}}(\vec{x}_i-\vec{x}_\text{c})$ from \cref{eq:r}.

\item[(b)] \emph{Large signals}.
For a large number of particles the distribution of the signal is modeled with a Gaussian approximation $\mathcal{N}(S_i;S(r_i),\sigma_i)$ which is used for large signals $S_i\geqslant S_\text{stat}$ and where $\sigma_i=\sigma_{S(r_i)}$ according to the \cref{eq:sigma_S}.

\item[(c)] \emph{Unrecoverable saturated signals}.
For saturated signals that fail the signal recovery procedure, mentioned in \cref{s:saturation}, $S_i$ is treated only as a lower limit to the actual size of the signal.
The Gaussian function $\mathcal{N}$ for large signals is integrated over all possible values larger than $S_i$ to get an estimate $F_\text{sat}(S_i,S(r_i))$ of the probability of detecting a signal larger than $S_i$, but only when no signal recovery is applied.

\item[(d)] \emph{Non-triggered stations}.
To trigger, a station has to acquire a certain amount of signal, a process that can be modeled with a trigger probability $p_\text{trig}(S(r_i))$, which is derived from data~\cite{Abraham:2010zz}.
The complementary probability for non-triggered stations is thus modeled as $F_\text{non}(S(r_i))=1-p_\text{trig}(S(r_i))$.

\end{itemize}
This results in a likelihood function
\begin{equation}
\mathcal{L} =
  \prod_i^{N_\text{small}} f_\text{Poi}(S_i,S(r_i)) \,
  \prod_i^{N_\text{large}} \mathcal{N}(S_i;S(r_i),\sigma_i) \,
  \prod_i^{N_\text{sat}} F_\text{sat}(S_i,S(r_i)) \,
  \prod_i^{N_\text{non}} F_\text{non}(S(r_i)).
\label{eq:offlineLikelihood}
\end{equation}

For the \emph{Herald} reconstruction, the LDF likelihood function is split into two parts based on whether stations have triggered or not.
While for the triggered stations, the signal $S_i$ is assumed to be a Gaussian fluctuation around an average value $S(r_i)$ given by the \cref{eq:loglog}, for the non-triggered stations, the relevant likelihood term is given, as in \emph{Observer}, by the complementary probability.
The \emph{Herald} reconstruction maximizes a likelihood function
\begin{equation}
\mathcal{L} =
  \prod\limits_i^{N_\text{sig}}
  \mathcal{N}(S_i; S(r_i), \sigma_i)
  \,
  \prod\limits_i^{N_\text{non}}
 F_\text{non}(S(r_i)).
  \label{eq:CDASLikelihood}
\end{equation}
For showers with triggered stations satisfying the lever-arm conditions and thus ensuring a strong constraint on the slope of the LDF in \cref{eq:loglog}, the parameter $\beta$ is always fitted in the \emph{Herald} reconstruction.

\begin{figure}[t]
\def\figh{0.35}
\centering
\includegraphics[height=\figh\columnwidth]{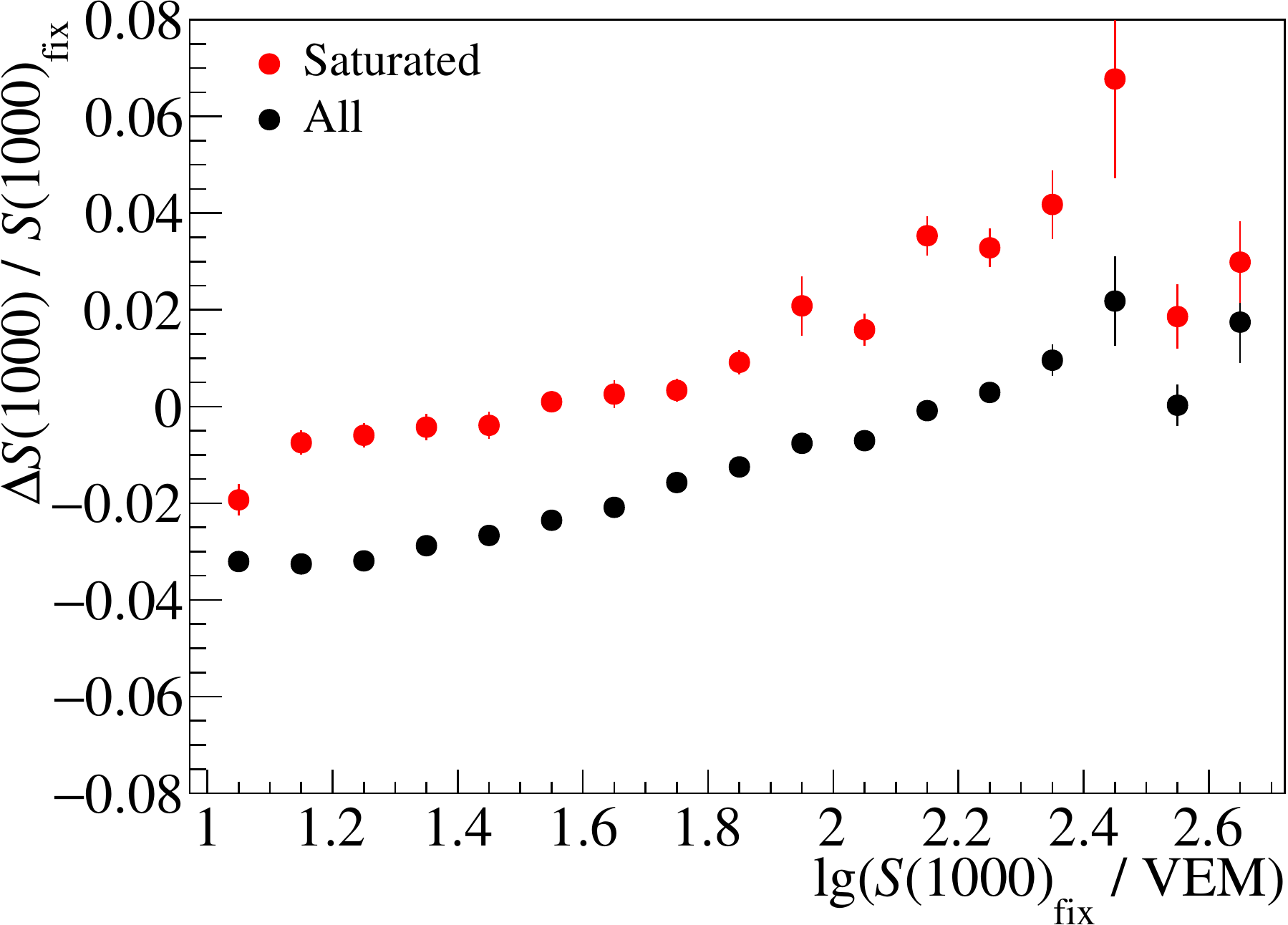}\hfill
\includegraphics[height=\figh\columnwidth]{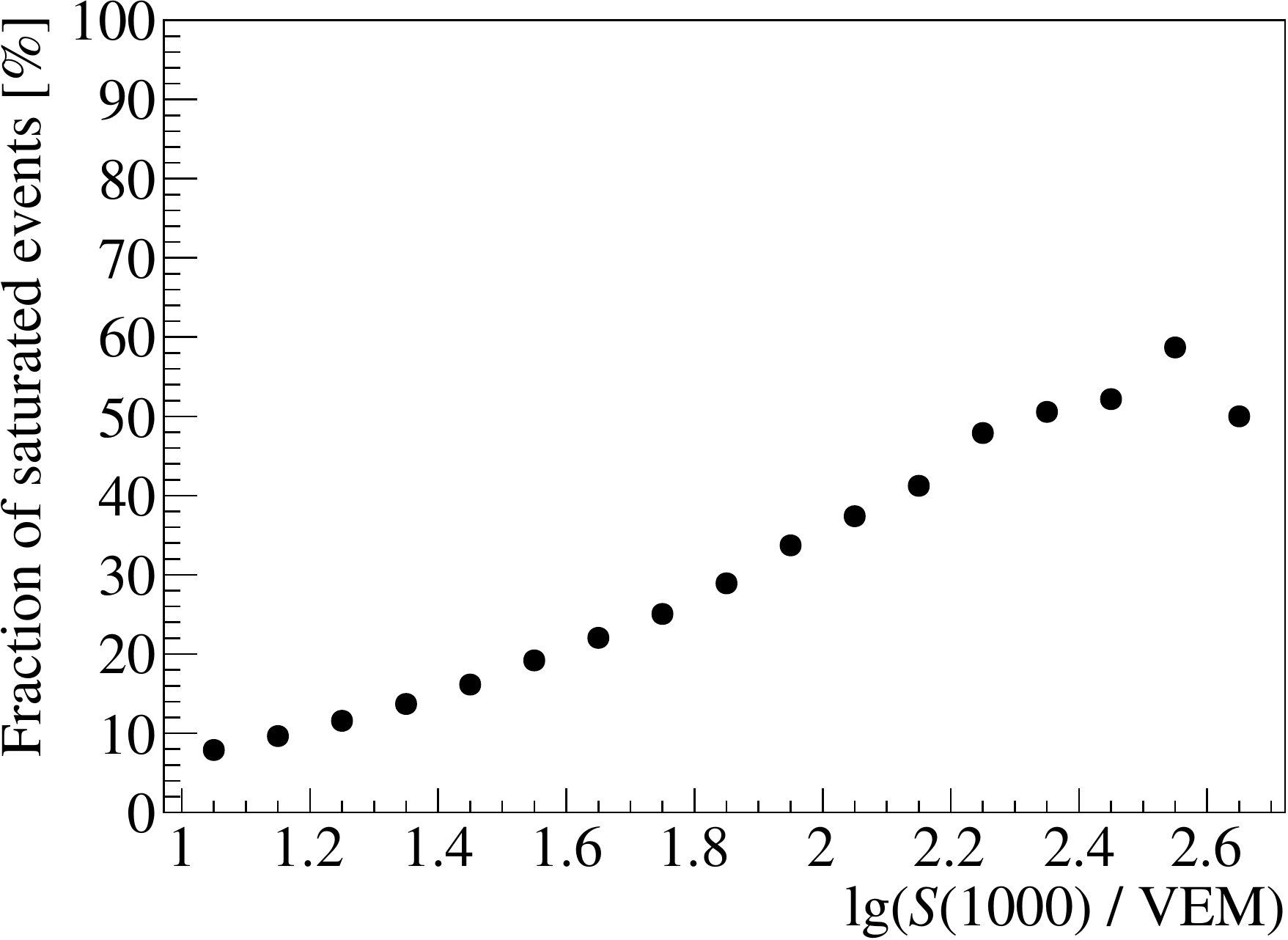}
\caption{\emph{Left}: Relative difference of the shower size due to the treatment of $\beta$ as a free or fixed parameter, obtained with the \emph{Observer} reconstruction where $\Delta S(1000)=S(1000)_\text{free}-S(1000)_\text{fix}$.
The relative difference is shown for all events in addition to events where a station saturated.
\emph{Right}: Fraction of saturated events as a function of the shower size.}
\label{fig:res1000}
\label{fig:SaturatedEvents}
\end{figure}

In the \emph{Observer} reconstruction, the parameter $\beta$ can be set as a free parameter for the same set of showers. 
A comparison of the shower sizes obtained with $\beta$ as a free or fixed parameter of the LDF in \cref{eq:nkg} is shown in \cref{fig:res1000}-left. 
The impact of freeing or fixing $\beta$ amounts to a less than 3\% (4\%) bias in the reconstructed value of $S(1000)$ for non-saturated (saturated) events and, therefore, fitting of $\beta$ is not required to obtain an unbiased estimate of the shower size~\cite{Hillas:1969zzb,Hillas:ICRC71}.
The fraction of all events which are saturated is ${\sim}$11\% but strongly depends on shower size as depicted in \cref{fig:SaturatedEvents}-right.
The fraction is constant to within 2\% across the full range of zenith angles.

Using the \emph{Herald} reconstruction, one final, global reconstruction step is performed to account for the coupling between the geometry and the LDF fits.
Depending on the number of degrees of freedom, any parameters which were previously fixed are at this stage allowed to vary.
The global log-likelihood function
\begin{equation}
-2\ln\mathcal{L}_\text{global} = \chi^2 - 2\ln\mathcal{L}
\end{equation}
is minimized, where the two terms are given in \cref{eq:chi2timing,eq:CDASLikelihood}.

\begin{figure}[t]
\def\figh{0.34}
\centering
\includegraphics[height=\figh\columnwidth]{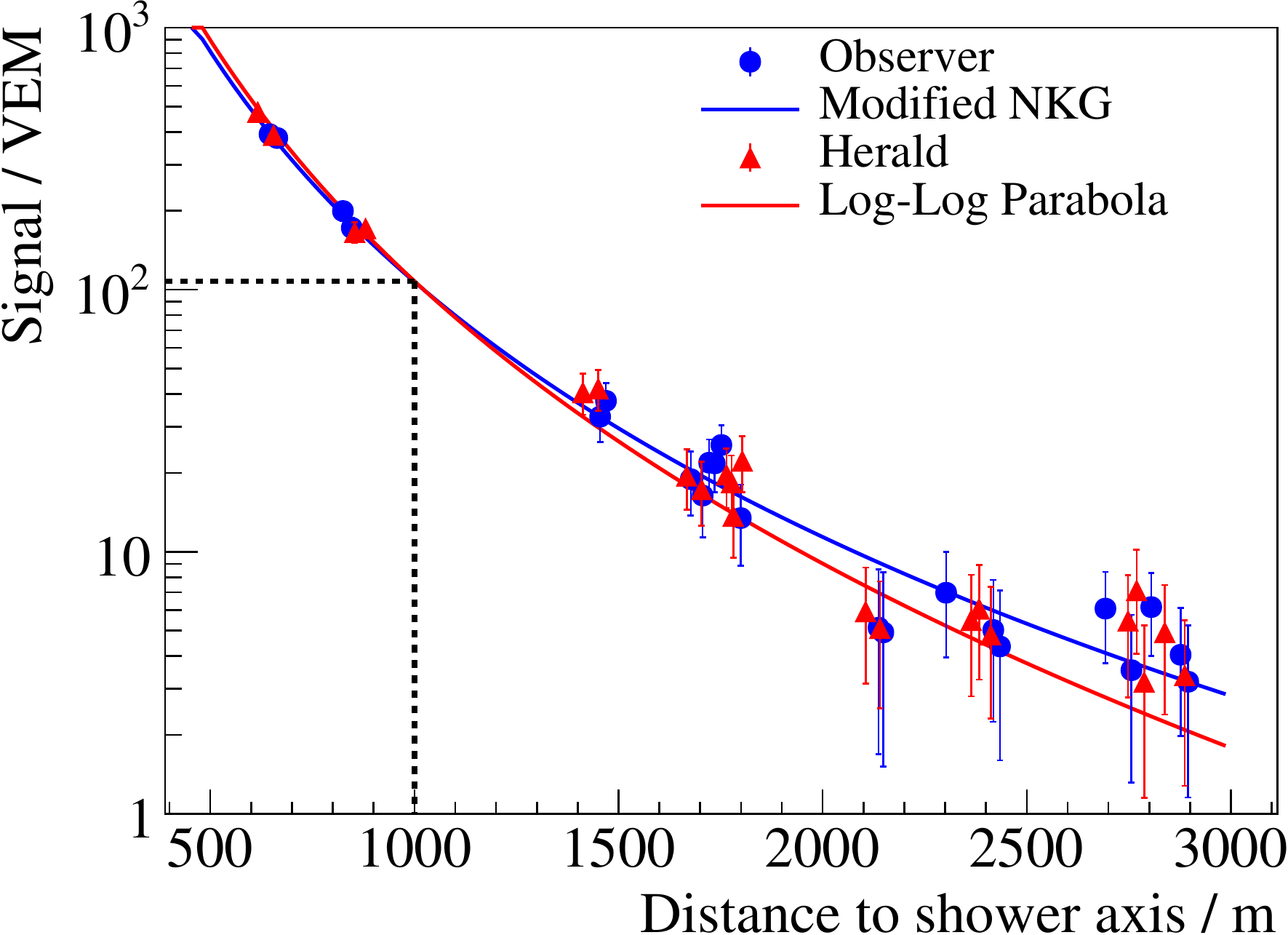}\hfill
\includegraphics[height=\figh\columnwidth]{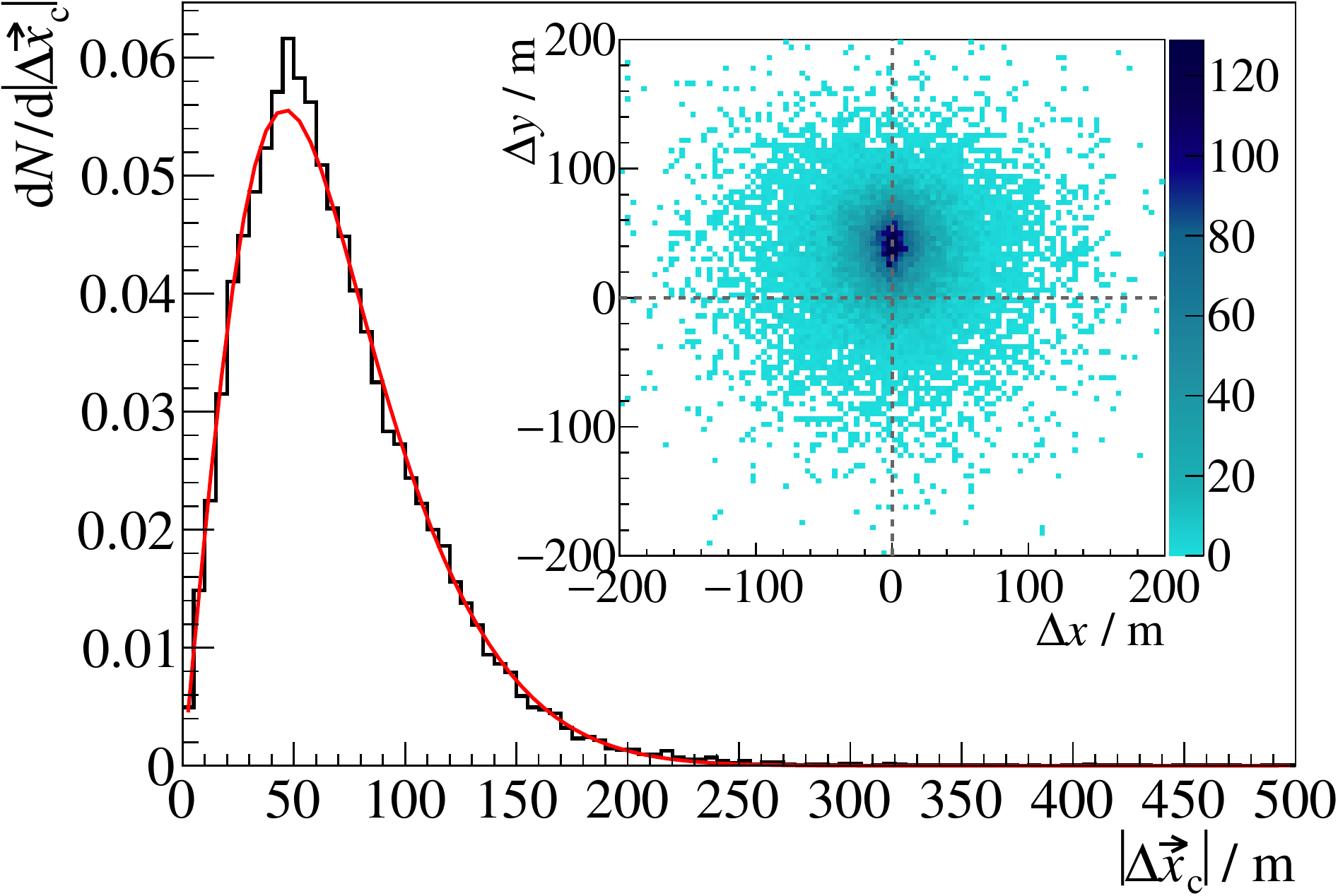}
\caption{\emph{Left:} The lateral distribution of signals from the example event in \cref{f:event} is plotted as a function of the radial distance of stations from the shower axis, shown for both \emph{Herald} (red triangles) and \emph{Observer} (blue circles) reconstructions, along with their respective LDFs, \cref{eq:loglog,eq:nkg}.
The reference signals at the radial distance of 1000\,m are indicated by the dashed line with values of $S(1000)=112$\,VEM and 107\,VEM for this particular event.
See \cref{f:event} for the image of the corresponding footprint.
\emph{Right:} Distribution of distances, $|\Delta\vec{x}_\text{c}|=|\vec{x}_\text{c}^\text{LLP}-\vec{x}_\text{c}^\text{NKG}|$ between the impact points reconstructed in two different ways: with a Log-Log Parabola (LLP) LDF in \emph{Herald} and a modified NKG in \emph{Observer}.
The distribution has been fitted with a Rayleigh function (red line) indicating that in 68\% of the events, the two reconstructed impact points are at a distance smaller than 92\,m.
\emph{Inset:} 2D distribution of $\Delta x=x_\text{LLP}-x_\text{NKG}$ and $\Delta y=y_\text{LLP}-y_\text{NKG}$.
These coordinates have been computed in the reference frame of the \emph{Observer} shower and are oriented so that the positive/negative $\Delta y$ axis is in the down/upstream direction.}
\label{f:ldf}
\label{f:core}
\end{figure}

The reconstructions described above successfully converge more than 99\% of the time.
The convergence failures are mostly due to low-multiplicity events formed by random coincidences which only by chance pass our selection procedures.
It is important to note that using the non-triggered stations adds a significant geometrical constraint to the LDF fit, in particular for low-multiplicity events.
Without consideration of the non-triggered stations, deviations of the shower size on the order of up to 8\% are observed.

An example of the LDF fits obtained with the \emph{Herald} and \emph{Observer} reconstructions is shown in \cref{f:ldf}-left.
In this example, the two LDF fits give very similar results for the estimation of the shower size $S(1000)$.
The differences between the two LDFs (defined in \cref{eq:loglog,eq:nkg}) are due to uncertainty in the shape of the underlying lateral distribution close to and far from the shower core.
These differences lead to a distribution of distances $|\Delta\vec{x}_\text{c}|$ between the shower impact points shown in \cref{f:core}-right.
For the two reconstructions, 68\% of the events exhibit a difference of less than 100\,m.
The systematic differences in the positions of the impact points (as shown in the inset of \cref{f:core}-right) are largely due to the asymmetry correction defined in \cref{eq:azimuthalAsymm}, which is only employed in the \emph{Herald} reconstruction.

\begin{figure}[t]
\def\figh{0.40}
\centering
\includegraphics[height=\figh\columnwidth]{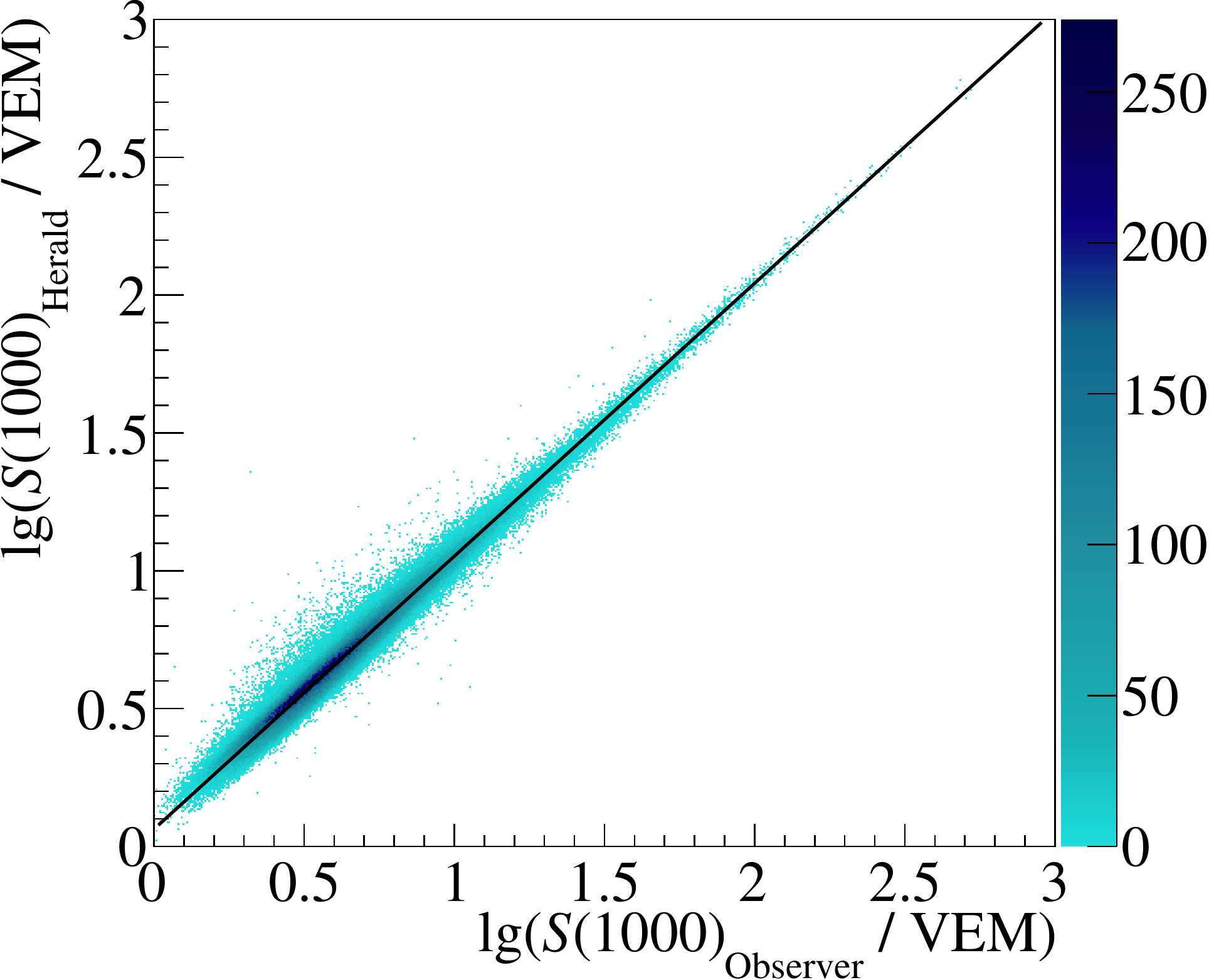}\hfill
\includegraphics[height=\figh\columnwidth]{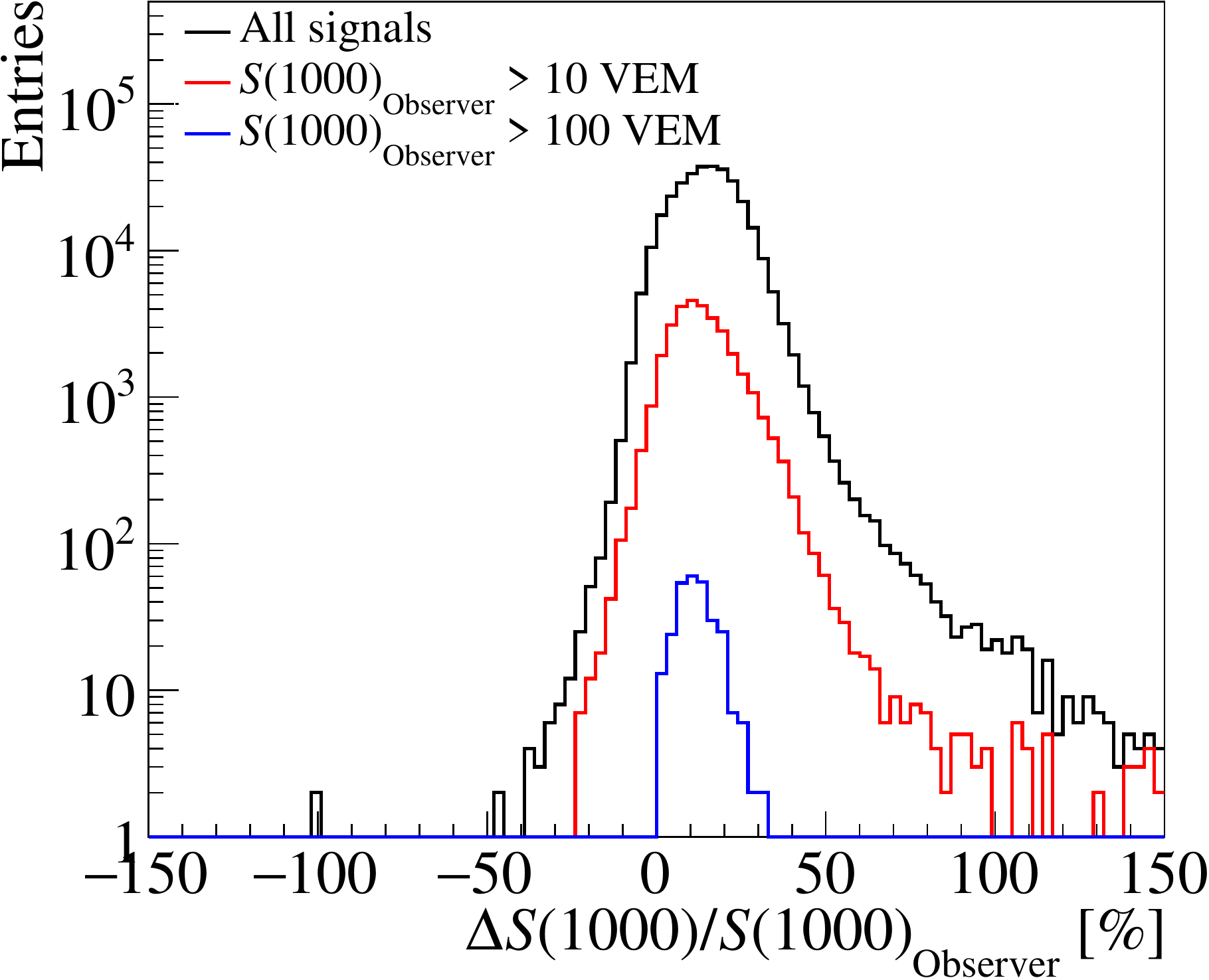}
\caption{\emph{Left:} Comparison of $S(1000)$ distributions for both reconstructions.
The distribution can be fitted with the following power law: $S(1000)_\text{Herald}/\text{VEM}=1.154(S(1000)_\text{Observer}/\text{VEM})^{0.9904}$ (in black line).
\emph{Right:} Relative difference $\Delta S(1000)/S(1000)_\text{Observer}$, where $\Delta S(1000)=S(1000)_\text{Herald}-S(1000)_\text{Observer}$, for all events and also only those with $S(1000)$ greater than 10 or 100\,VEM.
The corresponding means (standard deviations) of the distributions are 15.4\% (10.8\%), 11.2\% (9.2\%), and 10.9\% (5.3\%), respectively.}
\label{f:s1000}
\end{figure}

In \cref{f:s1000}, a similar comparison is shown for the shower-size parameter $S(1000)$.
The skewness of the distributions seen in \cref{f:s1000}-right is a consequence of a zenith-angle dependence of the systematic difference between the two reconstructions of $S(1000)$.
These systematic differences are corrected for in the angular portion of the calibration procedure, which is described in \cref{s:energy_estimator}.
The fit of $S(1000)_\text{Herald}$ vs.\ $S(1000)_\text{Observer}$ shows that the correlation between the two reconstructions follows a simple power law.
Deviations from this power law do not exceed 3\% at any point across the range of shower sizes.
Note that as long as the correlation follows a power law, the energy calibration procedure summarized in \cref{s:energy_estimator} will exactly reconcile any differences in the shower size estimates of the two reconstructions.

\section{Angular resolution}
\label{s:angularuncer}

The angular accuracy of the reconstruction of the SD events is mostly driven by the multiplicity of triggering stations and the precision of determination of the time at which the shower front arrived at each station.
The latter is governed by an interplay between the number of particles, their time distribution in the shower front, and the area of the detectors.
The small jitter of ${\sim}10$\,ns in the GPS timing system~\cite{Allison:2005vj} induces an angular uncertainty\footnote{~Increasing the jitter in the GPS timing to 25\,ns (100\,ns) would induce an effect ${\sim}0.2^\circ$ (${\sim}1.5^\circ$) for the smallest and ${<}0.1^\circ$ (${\sim}0.5^\circ$) for the largest shower sizes.} of less than ${\sim}0.1^\circ$.
The uncertainty introduced by the 40\,MHz sampling of the signals is of the same order of magnitude.

Deriving the angular resolution of cosmic-ray experiments usually relies on the use of simulations where the reconstructed arrival direction is compared with the injected one. 
We also present such a simulation-based derivation here but additionally show results of measurement-driven studies of the angular resolution.
The angular resolution is obtained by comparing two arrival directions $\hat{a}_1$ and $\hat{a}_2$, which are either (a) reconstructed and true directions in the case of simulations, (b) two arrival directions reconstructed by the same procedure and thus with the same resolution, or (c) two arrival directions obtained by two different reconstruction methods, which may have different resolutions.
The angular difference $\eta$ is obtained from $\sin\eta=|\hat{a}_1\times\hat{a}_2|$.
For $\eta\ll1$ and in a normal approximation with one-dimensional variance $\sigma^2$, $\eta$ follows a distribution $\mathrm{d}N/\mathrm{d}\eta=\mathcal{R}(\eta;\sigma)$ for the case of (a) and $\mathrm{d}N/\mathrm{d}\eta=\mathcal{R}(\eta;\sqrt{2}\sigma)$ for the case of (b), where $\mathcal{R}(r;\sigma)=(r/\sigma^2)\exp(-r^2/2\sigma^2)$ is the Rayleigh distribution.
We define the angular resolution ($\text{AR}$) as the angle at which the value of the cumulative distribution of $\eta$ reaches 68.3\%, this being in turn equivalent to $\text{AR}\approx1.52\sigma$ or $1.52\sigma/\sqrt{2}$ for the two cases, respectively~\cite{Bonifazi:2005ns}.
For the case (c) we assume that the two measurements $\sigma_1$ and $\sigma_2$ are uncorrelated and thus can be added in quadrature, i.e.\ $\sigma_\text{tot}^2=\sigma_1^2+\sigma_2^2$.

Below, three different approaches are explored to derive the angular resolution from measurements only, followed by a comparison of the results with a derivation from simulations\footnote{~Results from the \emph{Observer} framework are shown for these analyses and are compatible with those from \emph{Herald}.}.

\subsection{Derivation from measurements}

\subsubsection{Hybrid data}
\label{s:goldenhybriddata}

In addition to the independent triggering systems of the FD and SD detectors, a hybrid trigger~\cite{Abreu:2010aa} has been designed requiring a coincidence of at least one SD station and an FD telescope.
A sub-sample of these hybrid events can be fully reconstructed by both the FD hybrid reconstruction~\cite{Abraham:2009pm} and, independently, by the SD reconstruction.
These events are used, among other things, for the calibration of the SD energy estimator with the calorimetric measurement performed by the FD~\cite{spectrum_prd_2020}.
From the direct comparison of the arrival directions of the two reconstructions, $\hat{a}_\text{SD}$ and $\hat{a}_\text{hyb}$, the angular resolution $\text{AR}_\text{SD:hyb}$ is extracted from the angular difference $\eta_\text{SD:hyb}$ defined by $\sin\eta_\text{SD:hyb}=|\hat{a}_\text{SD}\times\hat{a}_\text{hyb}|$.
Given the essential independence of the SD and hybrid reconstructions, the SD angular resolution $\text{AR}_\text{SD}$ can be derived as $\text{AR}_\text{SD}^2=\text{AR}_\text{SD:hyb}^2-\text{AR}_\text{hyb}^2$, where $\text{AR}_\text{hyb}$ is the angular resolution of the hybrid events only.
With criteria to ensure the quality of the hybrid reconstruction, 29\,344 events are selected.
The details of the criteria applied can be found in~\cite{Bonifazi:2005ns,Dawson:2007di}.

Moreover, 1049 events are seen by at least two telescopes belonging to two different FD sites.
For these \emph{stereo} hybrid events, a separate\footnote{~Separate, but not completely independent since they share the same SD station used for the timing information.} hybrid reconstruction is performed for each FD site.
Computing the angular difference $\eta_\text{hyb}$ between these arrival directions $\hat{a}^\text{hyb}_1$ and $\hat{a}^\text{hyb}_2$, the resolution of the hybrid reconstruction of this particular selection of stereo events is estimated to be $\text{AR}_\text{hyb}=(1.07\pm0.05)^\circ$ assuming that resolutions for $\hat{a}^\text{hyb}_1$ and $\hat{a}^\text{hyb}_2$ are the same\footnote{~Note that stereo events have different geometries than standard hybrid (mono) events (the latter being on average closer to the telescopes) and therefore this number does not reflect the average resolution of the hybrid reconstruction which was estimated to be $0.5^\circ$ in~\cite{Bonifazi:2005ns}.}.
With $\text{AR}_\text{hyb}$ at hand, $\text{AR}_\text{SD}$ can be estimated as given above.
The results are shown in \cref{fig:AR} (squares) as a function of the zenith angle and the shower size.

\begin{figure}[t]
\centering
\includegraphics[width=0.75\columnwidth]{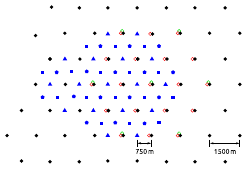}
\caption{A map of the high-density sector of the SD array which is in the North-West corner of the Observatory (see \cref{f:event}).
Regular stations of the 1500\,m grid are indicated with black markers.
Stations of the 750\,m grid are indicated with blue markers, where the stations forming three additional 1500\,m sub-grids are indicated with triangular, square, and pentagonal markers.
Twin and triplet stations in multiplets are indicated with red and green circles, respectively.}
\label{fig:InfillMap}
\end{figure}

\subsubsection{Super-hexagon of multiplets}
\label{s:superhexagonangularresoltion}

Located in the denser sector of the SD array, 12 twins and 7 triplets of stations (two or three stations separated by ${\sim}11$\,m) constitute the \emph{super-hexagon} of the Observatory, as schematically depicted in \cref{fig:InfillMap} (red and green circles).
The twins (or triplets) of the super-hexagon can be randomly divided to form two independent arrays and thereby enable two independent reconstructions of events that land in this region and trigger at least three twins.
All other stations outside of the super-hexagon are not included in this specialized event reconstruction.
Note that for showers of the same size, the multiplicity of stations used in this reconstruction can be, relative to the normal reconstruction, greatly reduced given the limited size of the super-hexagon.

Due to the small number of pairs available for this study, a selection on the position of the reconstructed core is applied. 
Only 913 reconstructed events\footnote{~Requirement of having the same number of stations in each twin set reduces the number of events to 301, nevertheless, the AR results remain statistically compatible.}, for which the impact point lies within a seed triangle constituted by twins only, are kept for the analysis. 
The estimate of the angular resolution $\text{AR}_\text{SD}$ is obtained from the distribution of the angular difference $\eta_\text{tw}$, defined as $\sin\eta_\text{tw}=|\hat{a}^\text{tw}_1\times\hat{a}^\text{tw}_2|$.
The results of this procedure are shown in \cref{fig:AR}.

\subsubsection{Sub-arrays}

The stations comprising the 750\,m array can be split into four small sub-arrays with the usual 1500\,m spacing, as illustrated in \cref{fig:InfillMap} (blue symbols).
Respectively, 841, 516, and 392 events are independently reconstructed two, three, and four times, whereby requiring that the impact point of each reconstruction lies inside the corresponding seed triangle. 

For a given event with several successful reconstructions, the arrival directions are compared in pairs, i.e.\ by computing the angular difference $\eta_{\text{sub}_{ij}}$ from $\sin\eta_{\text{sub}_{ij}}=|\hat{a}^\text{sub}_i\times\hat{a}^\text{sub}_j|$.
Here, we assume that all $\eta_{\text{sub}_{ij}}$ are independent, obtaining 4741 angles for the calculation of the estimate of angular resolution $\text{AR}_\text{SD}$.
The results are reported in \cref{fig:AR} (circles).

\subsection{Estimation using simulations}

Using simulations of extensive air-showers, the angular resolution is obtained as a function of zenith angle $\theta$ and the shower size $S(1000)$.
Showers initiated by a proton or an iron primary and developing according to the EPOS-LHC~\cite{Pierog:2013ria} or QGSJet-II.04~\cite{Ostapchenko:2010vb} model of hadronic interactions are simulated with energies between $10^{18.5}$ and $10^{20}$\,eV and zenith angles between $0^\circ$ and $60^\circ$.
An SD simulation and event reconstruction is performed 10 times for each shower simulation, where the impact position on the SD array is randomly chosen according to a uniform probability distribution across the complete array.

\begin{figure}[t]
\def\figh{0.37}
\centering
\includegraphics[height=\figh\columnwidth]{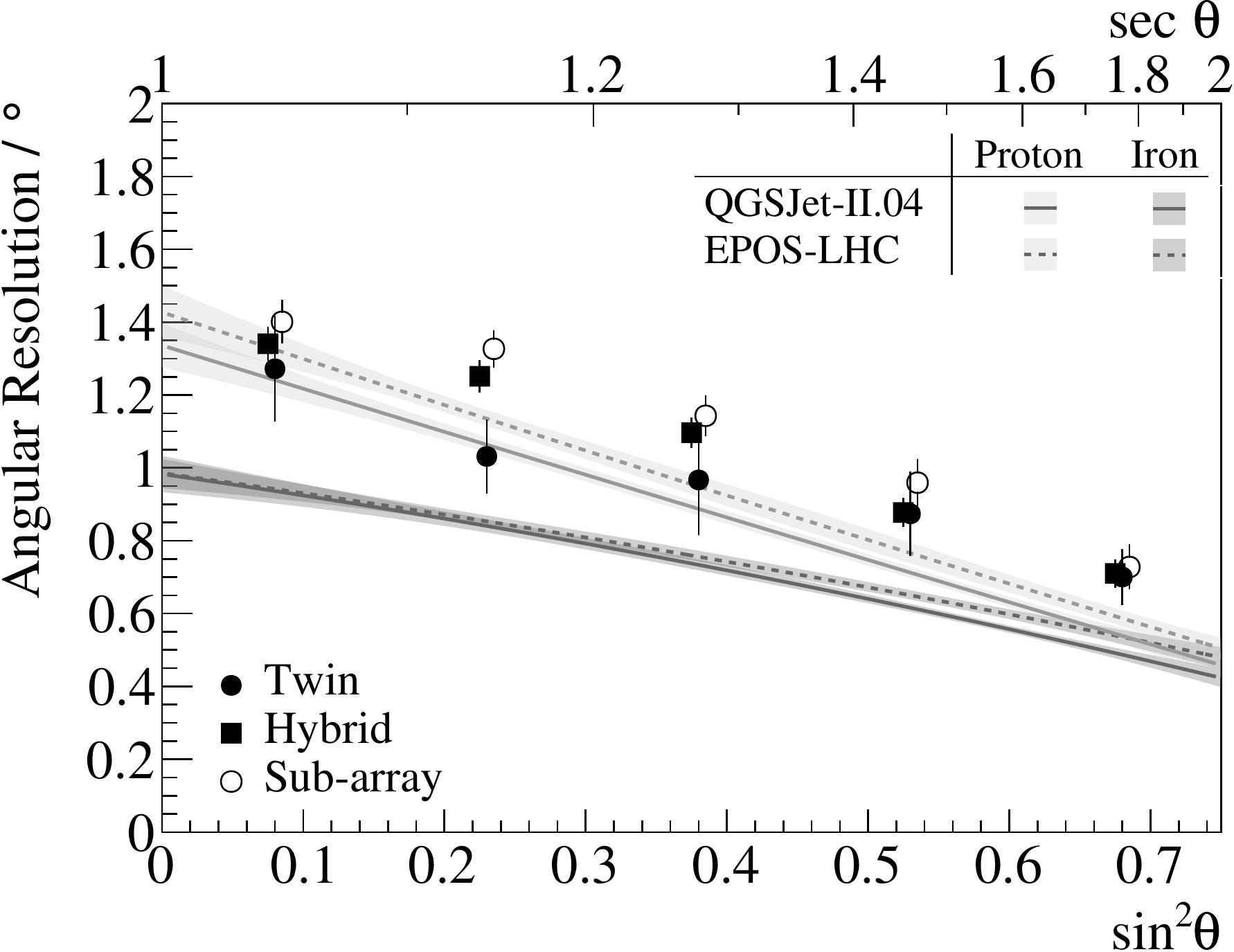}\hfill
\includegraphics[height=\figh\columnwidth]{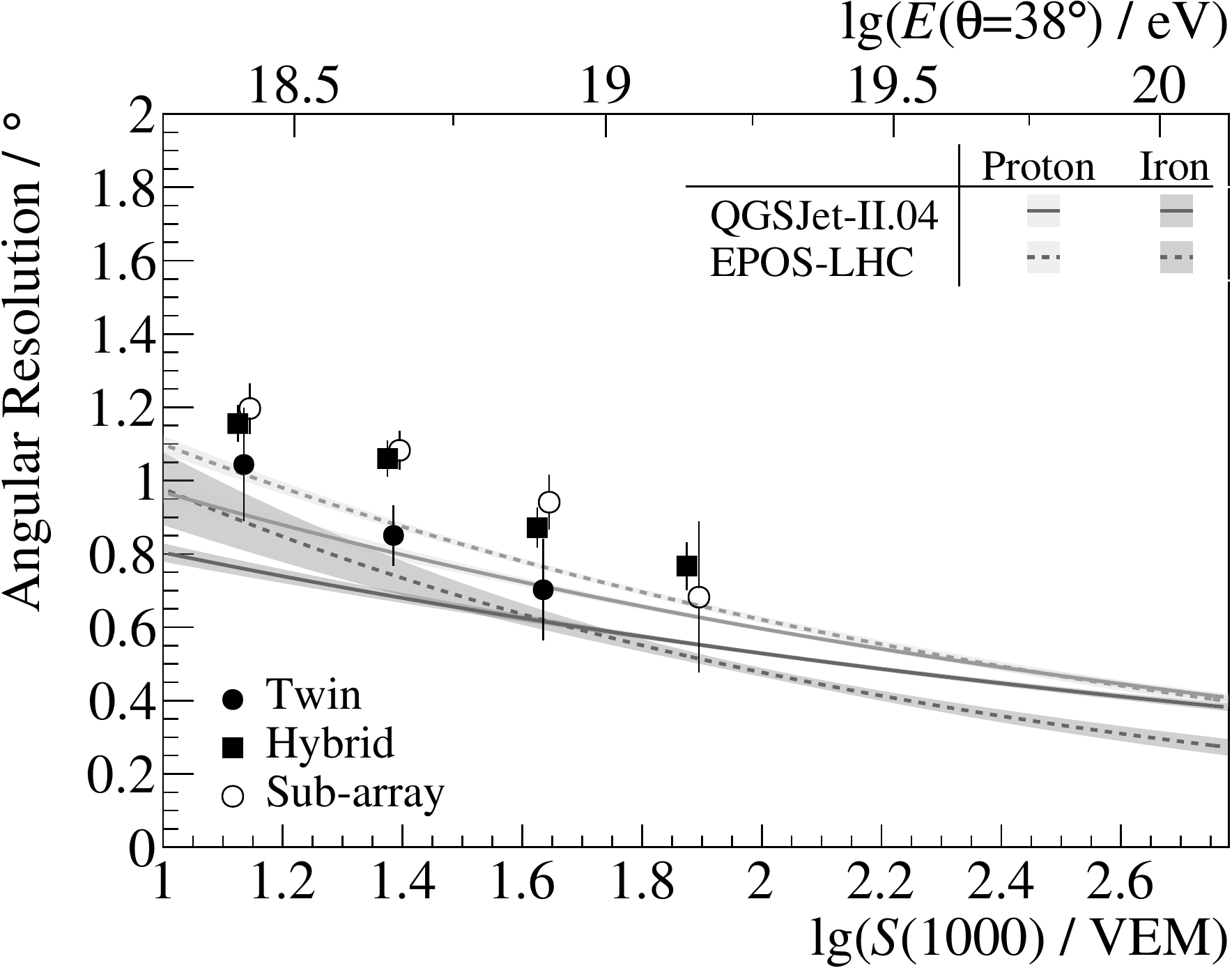}
\caption{Angular resolution of the reconstructed arrival direction as derived from a study of the hybrid events (filled black squares), a study of events with at least three triggered twins (filled black circles), and a study of three sub-arrays (open circles).
The lines report the results from the study with simulations produced with the hadronic model QGSJet-II.04 (full lines) and EPOS-LHC (dashed lines) for iron (dark gray lines) and proton (light gray lines) primaries.
The angular resolution of the reconstructed arrival direction is shown as a function of zenith angle (left) and the shower size (right).}
\label{fig:AR}
\end{figure}

The reconstructed arrival directions $\hat{a}_\text{rec}$ are compared with the true simulated axis $\hat{a}$ of the shower by computing the angular difference $\eta$ from $\sin\eta=|\hat{a}\times\hat{a}_\text{rec}|$ and estimating the AR from its distribution.
For plotting purposes, the results are empirically approximated to a good degree as a second-order polynomial in $\sin^2\theta$ and as an exponential function in $\lg(S(1000)/\text{VEM})$.
The results are reported in \cref{fig:AR} (lines and bands).

\subsection{Comparison of results}

The AR, as derived from both measured and simulated data, is shown to improve with increasing zenith angle and shower size.
The improvement for inclined events is attributed to the increase in the number of stations participating in an event due to the foreshortening of the array.

The slight discrepancies between the AR obtained from the measured and simulated data can be explained by limitations in the reconstruction of measurements in these specialized studies; considering that only a small number of SD stations were available for the twin and sub-array studies, the typical multiplicities of stations in events decreased and with them, the angular accuracy of the performed reconstructions.
Finally, computing the SD angular resolution from events reconstructed by both the FD and the SD in a data-driven way is limited by the knowledge of the angular resolution of the hybrid reconstruction $\text{AR}_\text{hyb}$. 
There are fewer high-quality events measured by the FD due to its limited duty cycle and the selection criteria applied.
Moreover, in order to derive $\text{AR}_\text{hyb}$ and then $\text{AR}_\text{SD}$, we distinguish between the FD reconstructions where the telescope was closer or further from the axis of the shower.
The geometrical reconstruction of the latter is thus expected to be worse, which would lead to an overestimation of the accuracy of the SD.

It is known that there are some deficiencies in the simulations of air showers, particularly with respect to the magnitude of the muonic component~\cite{Aab:2016hkv}.
Additionally, while simulations of the detector response have been thoroughly validated~\cite{Aab:2020wsn}, it is still possible that some experimental effects are not reproduced.
Performing comparisons of resolution estimates between simulated and measured data in terms of shower size mitigates the impact of the shortcomings of the simulations to the first-order.
The actual angular resolution of the SD is expected to lie somewhere between the measured data points and the simulated lines.
The angular resolution improves with both increasing zenith angle and shower size but is always better than $1.4^\circ$ and even approaches $0.7^\circ$ for the largest and most inclined showers.

It is also important to note that the differences between the results obtained from the \emph{Herald} and \emph{Observer} reconstructions are smaller than the angular resolution of the SD.
It is also worth mentioning that the angular differences observed in \cref{f:timing_angle}-right are primarily due to the correction of the signal asymmetry present in the \emph{Herald} reconstruction only.
These asymmetry corrections then induce a shift of the impact point of the shower core, as observed in \cref{f:core}-right, which is, in turn, translated into the angular differences observed.
The resolution of the impact point depends on the intrinsic properties of the shower and varies from ${\sim}100$\,m at the lowest energies to ${\sim}50$\,m at the highest energies.
The core resolution (measured in the ground plane) worsens with increasing inclination, which can be attributed to the projection effect.

\section{Uncertainties in \boldmath $S(1000)$}
\label{s:s1000uncer}

The uncertainty of the shower size estimator $S(1000)$ consists of statistical and systematic contributions.
While the statistical uncertainties of the reconstructed shower size, $\sigma_\text{stat}(S(1000))$, are directly related to the number of triggered stations and the uncertainties in their signals, the systematic uncertainties, $\sigma_\text{syst}(S(1000))$, arise from the lack of knowledge of the true shape of the LDF. 

The uncertainties of $S(1000)$ are first estimated through propagation of statistical and systematic errors in the fit of the LDF.
These estimations are complemented by two data-driven methods for the lower shower-size regime, where sufficient data are available, and a comparison with uncertainty estimates derived from simulations.

\subsection{Fit uncertainties and model-dependent systematic errors}

The statistical uncertainties in $S(1000)$ are directly estimated during the fitting procedure.
The derivation of the systematic uncertainties requires a deeper study of the slope parameter $\beta$ from \cref{eq:loglog,eq:nkg}.
As a first step, we use the same subset of events already used in the determination of $\beta$ in \cref{s:ldf}, i.e.\ events for which $\beta$ can be reliably fitted.
The resulting standard deviation $\sigma_\beta$ of the difference between the fixed and free $\beta$ is parametrized as a function of $S(1000)$.
In a second step, the systematic uncertainties are computed by propagating $\sigma_\beta(S(1000))$ into the shower size estimator for all events.

\begin{figure}[t]
\def\figh{0.37}
\centering
\includegraphics[height=\figh\columnwidth]{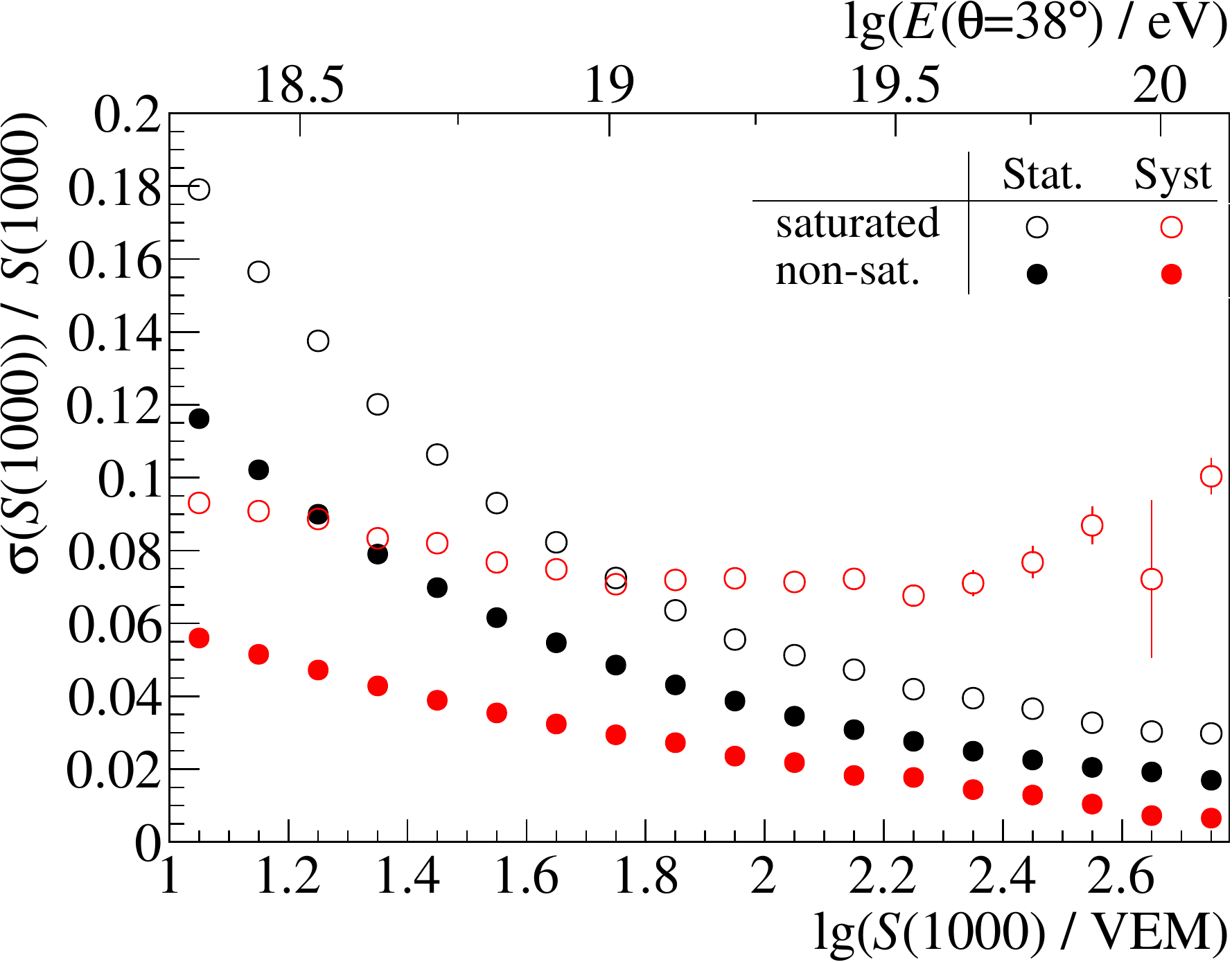}\hfill
\includegraphics[height=\figh\columnwidth]{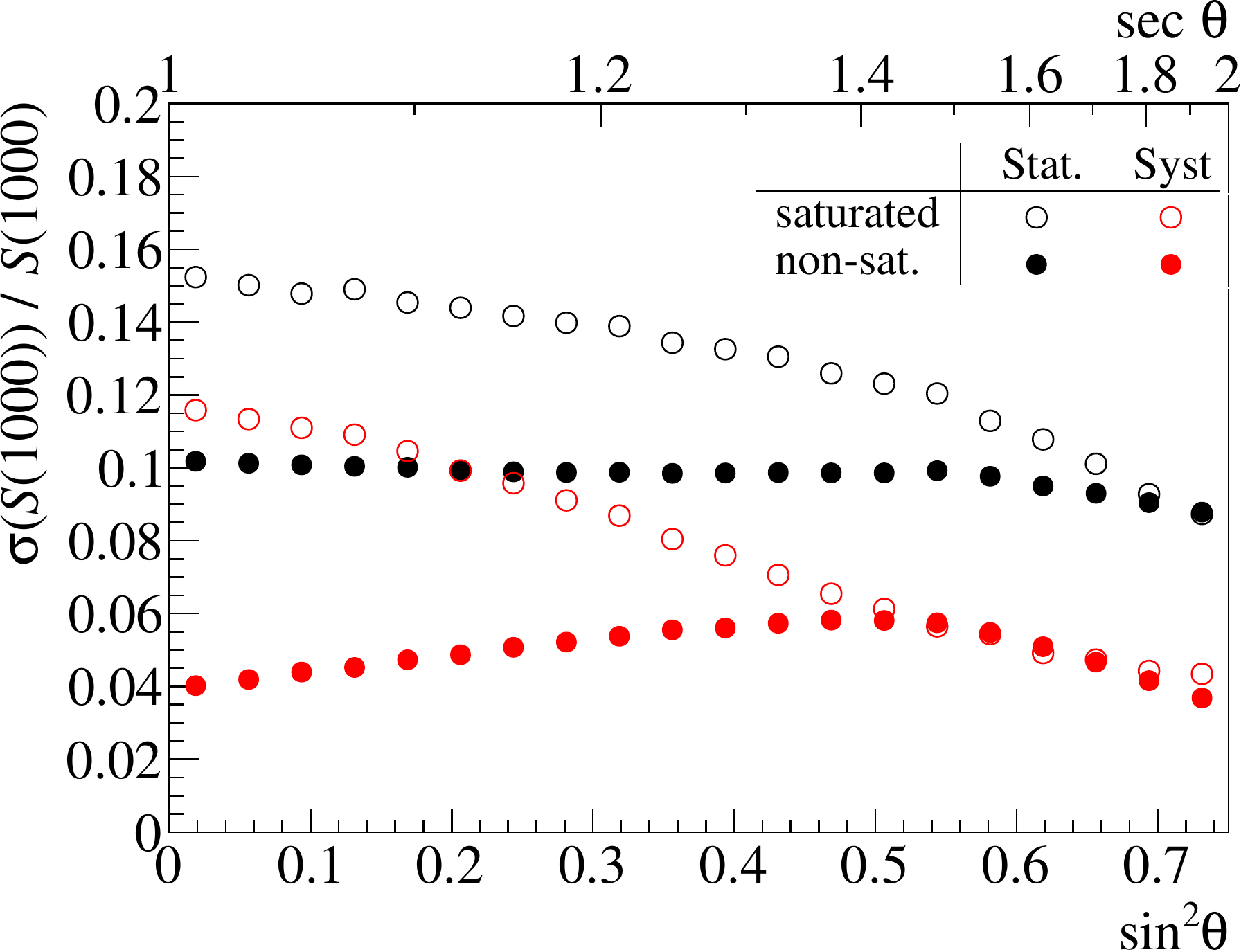}
\caption{Relative systematic (syst.) and statistical (stat.) uncertainty $\sigma(S(1000))/S(1000)$ of the shower-size reconstruction as a function of shower size $\lg(S(1000)/\text{VEM})$ (left) and $\sin^2\theta$ (right), obtained separately for saturated and non-saturated events.}
\label{fig:SystAndStat}
\label{fig:SystAndStat_theta}
\end{figure}

This approach is applied to the full data set of events recorded by the SD, while using the quality cuts described in the previous sections.
The average systematic uncertainties and statistical errors are reported in \cref{fig:SystAndStat} as a function of $\lg(S(1000)/\text{VEM})$ (left) and as a function of $\sin^2\theta$ (right) for saturated and non-saturated events.
In both cases, the statistical uncertainty dominates the total uncertainty for smaller shower sizes; however, for larger shower sizes, the systematic uncertainties become significantly larger in the case of saturated events.
This implies that an accurate description of the shape of the lateral distribution is of increasing relevance at the highest energies, where our estimates of the shower size may benefit from an improved knowledge thereof.
Note that the uncertainties in \cref{fig:SystAndStat_theta}-right are practically constant for the non-saturated events when considering all energies together.
On the other hand, the resolution for the saturated events improves with increasing zenith angle.
This is a consequence of the fact that the saturated station for most of the near-vertical saturated events is close to the shower core, resulting in a clustering of the six nearest stations around a distance of ${\sim}1500$\,m from the shower axis, which is not the case for more inclined showers.
At the highest energies, these clustered configurations become responsible for the systematic uncertainties being twice as high as the statistical uncertainties, and the recovery procedure of the saturated signals (\cref{s:saturation}) has an effect of only around 2\% on the shower size estimator.
A better knowledge of the shape of the lateral distribution is thus needed to reduce the uncertainties in $S(1000)$. 
Plans to improve this knowledge are discussed in \cref{s:outlook}.

\subsection{Direct measurements}

\subsubsection{Super-hexagon of multiplets}

As in the studies of the angular resolution discussed in \cref{s:superhexagonangularresoltion}, it is also possible to perform independent reconstructions of $S(1000)$ for events with at least three triggered twins located in the super-hexagon for which the quantity $\delta_\text{tw}=\Delta S(1000)_\text{tw}/\sqrt{2}\langle S(1000)\rangle_\text{tw}$ is computed, where $\Delta S(1000)_\text{tw}$ and $\langle S(1000)\rangle_\text{tw}$ are the difference and the average, respectively, of the two reconstructed shower sizes $S(1000)_1^\text{tw}$ and $S(1000)_2^\text{tw}$.
In each bin in $\lg(S(1000)/\text{VEM})$ or in $\sin^2\theta$, the standard deviation $\sigma(\delta_\text{tw})$ is calculated.
The standard deviations of these distributions are plotted in \cref{fig:CompSimuData} (as filled squares).

\subsubsection{Sub-arrays}

For events landing in the denser sector of the SD, a similar procedure as the one used for twin study is applied. 
Using the events reconstructed two, three, or four times with the stations from different sub-arrays, the quantity $\delta_\text{sub}=\Delta S(1000)_\text{sub}/\sqrt{2}\langle S(1000)\rangle_\text{sub}$ is derived, where $\Delta S(1000)_\text{sub}$ and $\langle S(1000)\rangle_\text{sub}$ are the difference and the average of the reconstructed shower sizes $S(1000)^\text{sub}_i$ and $S(1000)^\text{sub}_j$, respectively.
For each pair of sub-arrays $(i,j)$, the standard deviations $\sigma(\delta_\text{sub})_{ij}$ are calculated and then averaged over all pairs $(i,j)$ in different bins of $\lg(S(1000)/\text{VEM})$ and $\sin^2\theta$. 
The results are reported in \cref{fig:CompSimuData} (as open squares).

\subsection{Estimation using simulations}

A discrete set of simulations is used to determine the total uncertainty of the energy estimator $S(1000)$. 
Showers initiated by a proton or an iron primary and developing according to either the EPOS-LHC or the QGSJet-II.04 hadronic model are simulated with energies of $10^{18.5}$, $10^{19}$, and $10^{19.5}$\,eV and zenith angles of $0^\circ$, $12^\circ$, $22^\circ$, $32^\circ$, $38^\circ$, $48^\circ$, and $56^\circ$.
An SD simulation and event reconstruction is performed 10 times for each simulated shower. 

\begin{figure}[t]
\def\figh{0.35}
\centering
\includegraphics[height=\figh\columnwidth]{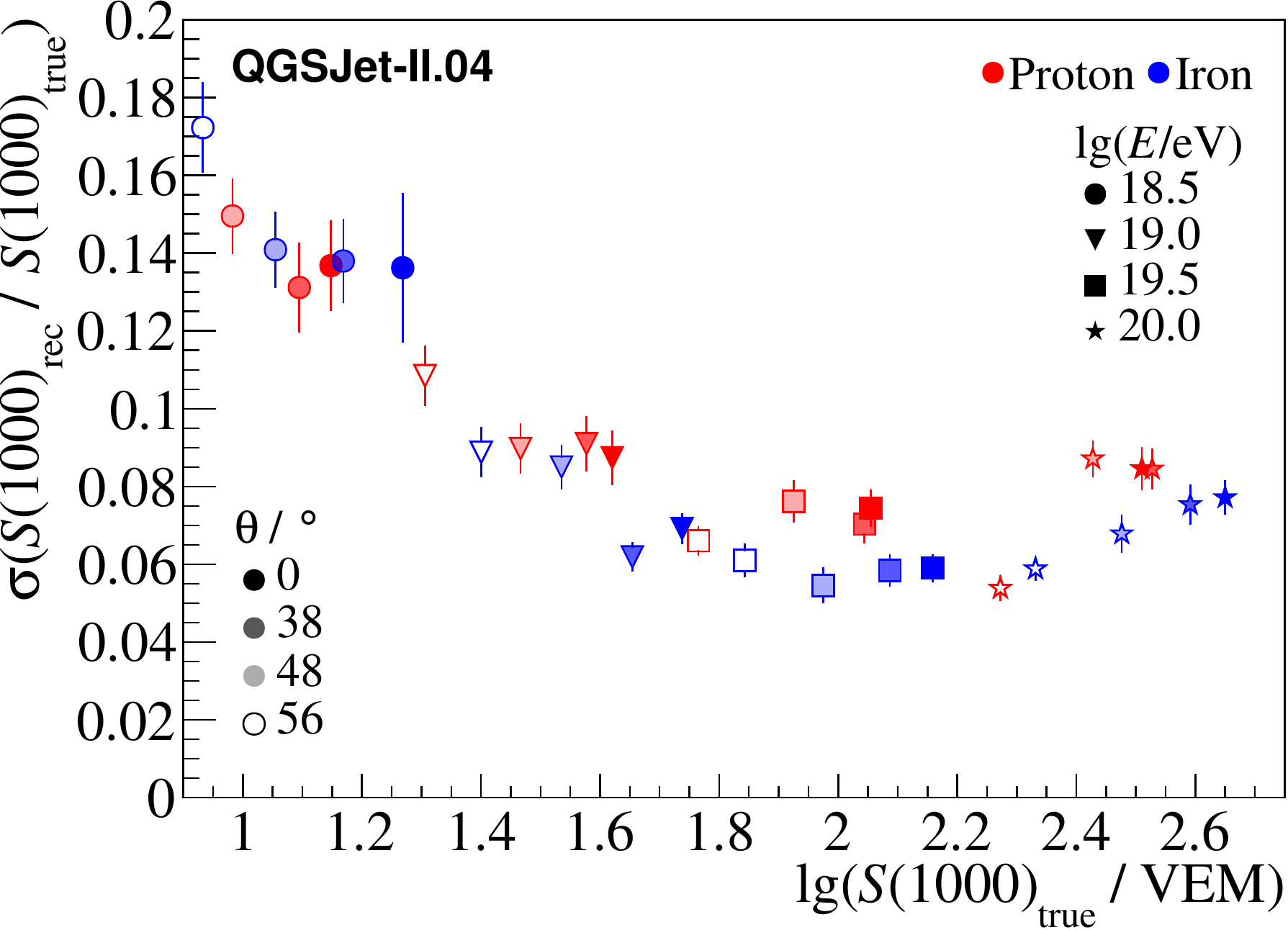}\hfill
\includegraphics[height=\figh\columnwidth]{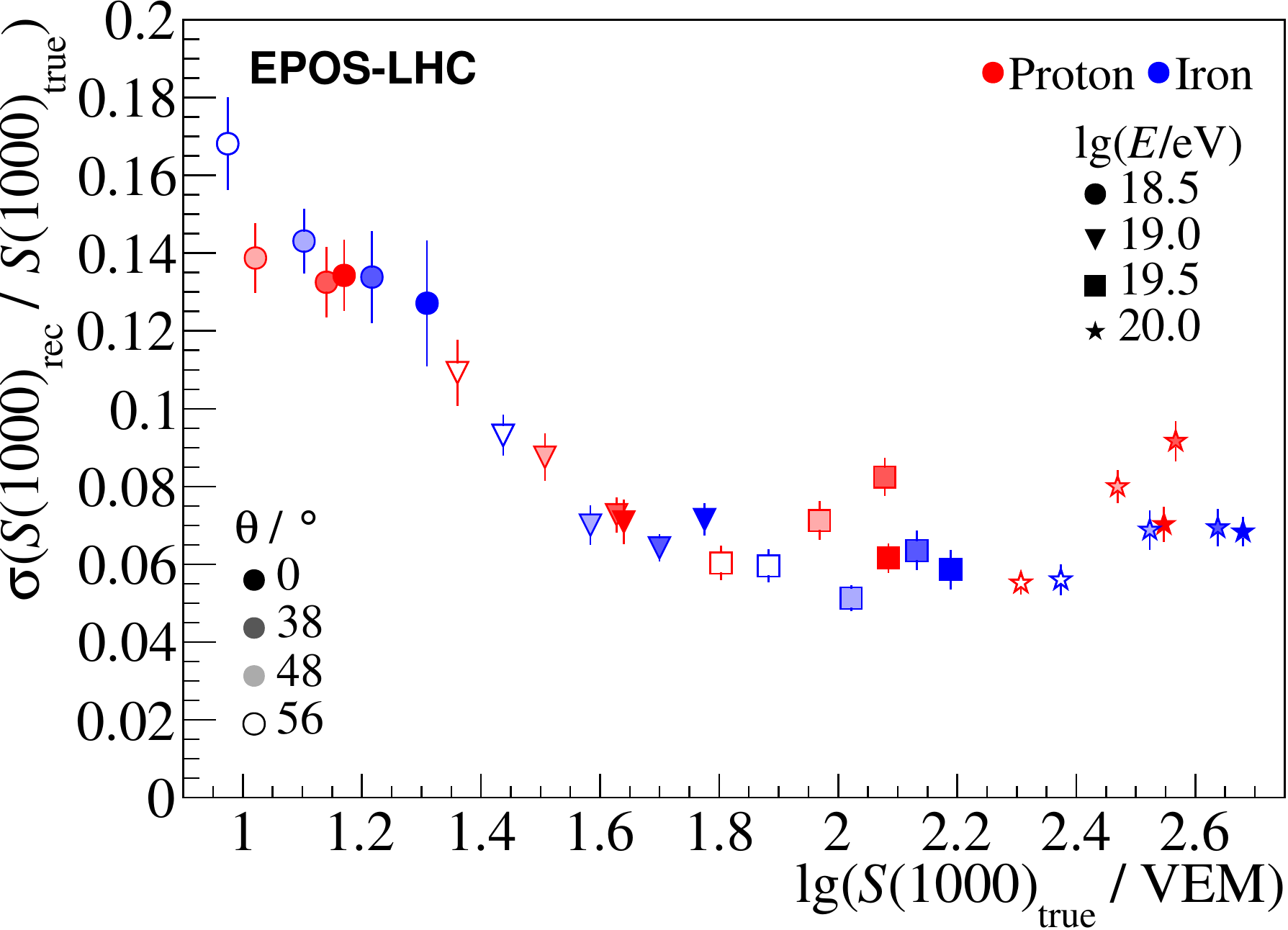}
\caption{Relative uncertainty of the reconstructed shower size $\sigma(S(1000)_\text{rec})/S(1000)_\text{true}$ as obtained from QGSJet-II.04 (left) and EPOS-LHC (right) simulations of proton (red) and iron (blue) primaries for several initial Monte-Carlo energies and zenith angles of $0^\circ$, $12^\circ$, $22^\circ$, $32^\circ$, $38^\circ$, $48^\circ$, and $56^\circ$ (indicated with increasing transparency of the markers).}
\label{fig:S1000Errors}
\end{figure}

For simulations, the true value of $S(1000)$, denoted as $S(1000)_\text{true}$, is determined by calculating the mean signal of 24 additional stations in a ring with a shower-plane radius of exactly 1000\,m.
The shower sizes $S(1000)_\text{true}$ and $S(1000)_\text{rec}$, where the latter is the value of the shower size reconstructed according to the procedure detailed in \cref{s:ldf}, are compared for different primaries and for different zenith angles by calculating the ratio $\delta_\text{sim}=\Delta S(1000)_\text{sim}/S(1000)_\text{true}$ where $\Delta S(1000)_\text{sim}=S(1000)_\text{rec}-S(1000)_\text{true}$.
The standard deviation $\sigma(\delta_\text{sim})$ of the relative differences is reported in \cref{fig:S1000Errors}.
The results show a total uncertainty in $S(1000)_\text{rec}$, which decreases from 18\% at the smallest shower sizes to nearly 7\% at the largest values of $S(1000)_\text{true}$.
At larger inclinations, the multiplicity of stations within a given distance from the shower axis increases, whereas the magnitudes of the signals decrease due to the larger attenuation of showers in the atmosphere.
These competing effects, which are shown in \cref{fig:S1000Errors}, result in the observed evolution of the standard deviation $\sigma(\delta_\text{sim})$ with zenith angle.

\subsection{Comparison of results}

\begin{figure}[t]
\def\figh{0.37}
\centering
\includegraphics[height=\figh\columnwidth]{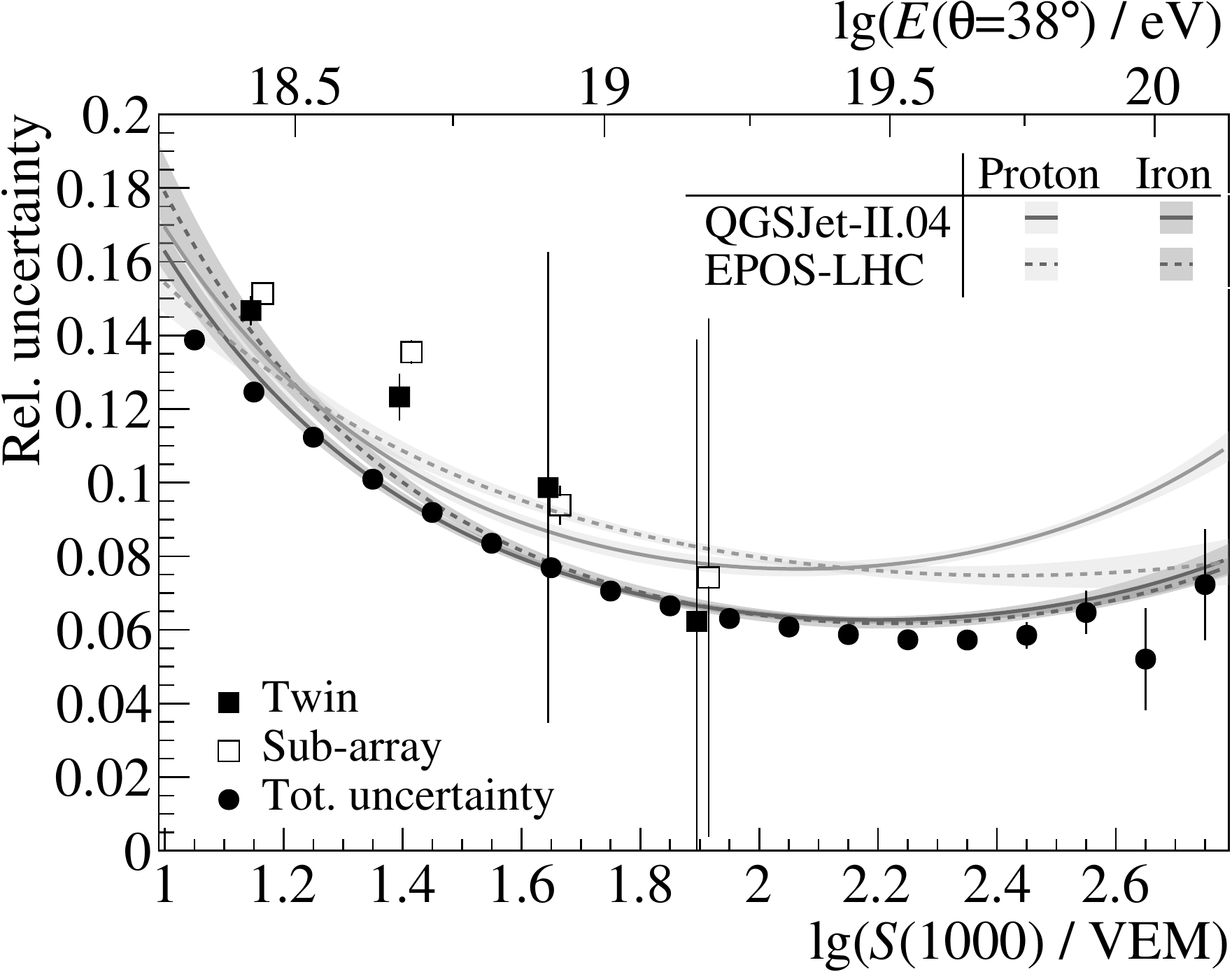}\hfill
\includegraphics[height=\figh\columnwidth]{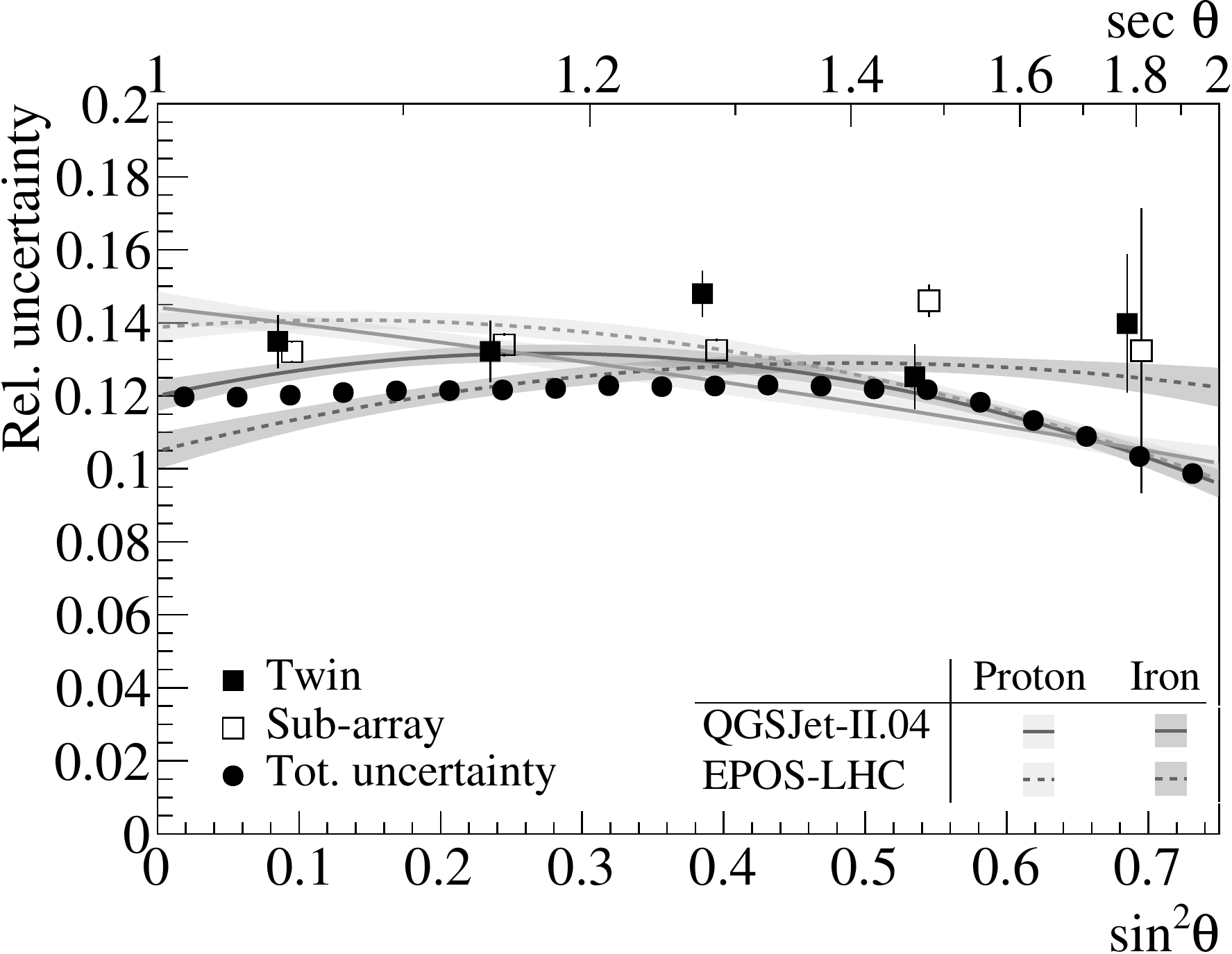}
\caption{Relative total uncertainty of the shower-size reconstruction $\sigma(S(1000))/S(1000)$ derived from a study of statistical and systematic uncertainties separately (black circles), a study of events with at least three triggered twins (black squares), and a study of three sub-grid arrays (black open squares). 
The lines report the results from the study with simulations produced with the hadronic models QGSJet-II.04 (full lines) and EPOS-LHC (dashed lines) for iron (dark gray lines) and proton (light gray lines) primaries.
The relative total uncertainty is shown as a function of the shower size (left) and zenith angle (right).}
\label{fig:CompSimuData}
\end{figure}

The statistical and systematic uncertainties of the reconstructed $S(1000)$ for all measured data are added in quadrature as $\sigma^2(S(1000)_\text{rec})=\sigma^2_\text{syst}(S(1000)_\text{rec})+\sigma^2_\text{stat}(S(1000)_\text{rec})$, where $\sigma_\text{syst}$ is from \cref{fig:res1000}-left.
The result is compared with both data-driven and simulation-based estimates in \cref{fig:CompSimuData}, where the results from simulations have been parameterized with simple polynomials.

While some discrepancies may be observed and can be attributed to the systematic errors associated with each method of estimation, all methods generally agree to within a few percentage points.
The slight worsening of the resolution from $\sim$6\% at $\lg(S(1000)/\text{VEM})\approx2.3$ to $\sim$8\% at the largest shower sizes is due to the fact that at a shower size of $S(1000)=200$\,VEM, 50\% of the events have a saturated station and this fraction increases with shower size.
As seen in \cref{fig:CompSimuData}-right, the resolution is relatively constant at 14\% (since small shower sizes dominate calculation of these estimates) over the zenith angle range with a slight decrease observed for the most inclined shower sizes.

The impact of the resolution in shower size on the uncertainties in the estimate of the primary energy is described in detail in~\cite{spectrum_prd_2020}.

\section{Towards an efficient estimate of the energy}
\label{s:energy_estimator}

In the previous sections, we presented the reconstruction of the arrival direction and the shower size from the timing and signal measurements of SD stations.
The estimation of the shower size is a first important step in the determination of the energy of the primary cosmic ray. 
The second step involves the correction of a number of biases to improve the energy resolution.
The shower size is further corrected for daily and yearly atmospheric variations, which change the effective atmospheric overburden of the array.
This correction has been recently improved~\cite{Aab:2017kgl} using eleven years of data from the Observatory.
Next, a bias resulting from (oppositely) charged shower particles experiencing deflection in the geomagnetic field is removed~\cite{Abreu:2011ki}.
This effect causes a dependence of the shower size on the azimuthal orientation of the shower with respect to the orientation of the local geomagnetic field.
The shower size is corrected with a factor $1+A_\text{gm}\,\cos^k\theta\,\sin^2\xi$, where $A_\text{gm}$ and $k$ are model parameters estimated from the Monte-Carlo data, and $\xi$ is the angle between the geomagnetic field and the shower axis.
The maximum value of this correction stays below 4\%.

The next steps are described in detail in \cite{spectrum_prd_2020}, but are briefly summarized here.
Using the constant intensity cut (CIC) method~\cite{Hersil:1961zz,GH:ICRC1977}, the measured shower size $S(1000)$ is converted to an angle-independent shower size $S_{38}$.
This is equivalent to the size of the shower had the primary particle arrived with a zenith angle of $38^\circ$ (median of all events), $S_{38}=S(1000)/f_\text{CIC}(\theta)$.
This correction accounts for the increasing slant depth of the array with increasing zenith angle.
Since the more inclined showers are sampled at a later shower age, the more-attenuated electromagnetic cascade leads to a smaller observed shower size.
With this procedure, we obtain the minimally biased, zenith-independent energy estimator $S_{38}$.
This can be directly calibrated~\cite{Dembinski:2015wqa} by the nearly-calorimetric energy measurement of the FD using hybrid events (see \cref{s:goldenhybriddata}).
The power-law calibration curve $E=A\,(S_{38}/\text{VEM})^B$, obtained with these \emph{hybrid} events, is then used to assign the energy to \emph{all} SD events.
All remaining lowest-order differences in reconstruction between the \emph{Herald} and \emph{Observer} frameworks are essentially removed by performing this energy calibration step independently for each.

\section{Conclusion and outlook}
\label{s:outlook}

We have presented the methods used for reconstructing the properties of the highest-energy cosmic rays from measurements of the surface-detector array of the Pierre Auger Observatory.
This description ranges from the station-level calibration procedure to the event-level reconstruction of the arrival direction and the shower size.
At the core of these methods is an emphasis on measurement-driven approaches.

At each step of the reconstruction procedure, whether it be modeling the curvature of the impinging shower front or fitting the lateral distribution of particles at the ground, a number of differing approaches may be developed and applied, each with its own assumptions.
Within the Pierre Auger Collaboration, exploration and validation of different assumptions and methods has been facilitated by the two reconstruction frameworks, \emph{Herald} and \emph{Observer}.
The existence of these independent frameworks has provided a source of cross-validation and has placed a spotlight onto the impact of different procedures on the reconstructed observables of the primary cosmic-rays throughout the development of our reconstruction procedures.
It has also resulted in the production of high-quality data sets of reconstructed events.
Historically, the \emph{Herald} data set has been used for anisotropy analyses and that of \emph{Observer} for the energy spectrum and mass composition studies, where one or the other was chosen out of practical reasons for publication.
Nonetheless, the consistency of physical results has been verified using both frameworks.
The angular resolution of the reconstructed arrival direction was shown to be of the order of $1^\circ$ and approaches $0.5^\circ$ for the largest shower sizes (i.e.\ above 200\,VEM).
The resolution in the reconstructed shower size, from which the energy of the primary is inferred, was demonstrated to improve from ${\sim}15\%$, for the smallest shower sizes down to ${\sim}6\%$ for the largest.
These values correspond to primaries of approximately $10^{18.5}$\,eV and above $10^{19.5}$\,eV, respectively (for more detailed interpretation in terms of primary energy, see~\cite{spectrum_prd_2020}).
At the highest energies, systematic uncertainties due to imperfect knowledge of the shape of the lateral distribution within a few hundred meters of the shower core become of the same order as statistical uncertainties, which will be addressed by the extension of the dynamic range of the WCDs through the installation of a smaller PMT~\cite{Castellina:2017ite}.

Although the reconstruction methods presented here were developed and applied to the 1500\,m array of the Observatory, scaled adaptations thereof are also successfully applied to the lower-energy extension of the surface detector array with a denser detector spacing of 750\,m~\cite{ThePierreAuger:2015rma}.
In both cases, the results are used for higher-level physics analyses, including studies of anisotropies in arrival directions~\cite{Aab:2017tyv,Aab:2018chp}, multimessenger~\cite{Aab:2019gra} and mass-composition studies~\cite{Aab:2016htd,Aab:2017cgk}, and the cosmic-ray energy spectrum~\cite{spectrum_prd_2020,spectrum_prl_2020,Coleman:2019rQ}.

Looking to the future, the surface detector of the Pierre Auger Observatory is currently undergoing a large-scale upgrade called AugerPrime~\cite{Aab:2016vlz,Castellina:2019huz}, with the placement of a 3.8\,m$^2$ scintillator atop each water-Cherenkov detector as one of the principal components.
Exploitation of the differing responses of the two detector types to the electromagnetic and muonic components of extensive air showers will permit further improvements to the event reconstruction algorithms described here and, most notably, add a primary-mass estimate to the set of reconstructed observables.

\acknowledgments
\input{acknowledgments}

\bibliographystyle{JHEP}
\bibliography{bibliography}{}

\begin{center}
\rule{0.1\columnwidth}{0.5pt}\,\raisebox{-0.5pt}{\rule{0.05\columnwidth}{1.5pt}}~\raisebox{-0.375ex}{\scriptsize$\bullet$}~\raisebox{-0.5pt}{\rule{0.05\columnwidth}{1.5pt}}\,\rule{0.1\columnwidth}{0.5pt}
\end{center}

\section*{The Pierre Auger Collaboration}
\input{latex_authorlist_authors}
{\footnotesize
\input{latex_authorlist_institutions}
}

\end{document}

%% file: acknowledgments.tex


\begin{sloppypar}
The successful installation, commissioning, and operation of the Pierre
Auger Observatory would not have been possible without the strong
commitment and effort from the technical and administrative staff in
Malarg\"ue. We are very grateful to the following agencies and
organizations for financial support:
\end{sloppypar}

\begin{sloppypar}
Argentina -- Comisi\'on Nacional de Energ\'\i{}a At\'omica; Agencia Nacional de
Promoci\'on Cient\'\i{}fica y Tecnol\'ogica (ANPCyT); Consejo Nacional de
Investigaciones Cient\'\i{}ficas y T\'ecnicas (CONICET); Gobierno de la
Provincia de Mendoza; Municipalidad de Malarg\"ue; NDM Holdings and Valle
Las Le\~nas; in gratitude for their continuing cooperation over land
access; Australia -- the Australian Research Council; Brazil -- Conselho
Nacional de Desenvolvimento Cient\'\i{}fico e Tecnol\'ogico (CNPq);
Financiadora de Estudos e Projetos (FINEP); Funda\c{c}\~ao de Amparo \`a
Pesquisa do Estado de Rio de Janeiro (FAPERJ); S\~ao Paulo Research
Foundation (FAPESP) Grants No.~2019/10151-2, No.~2010/07359-6 and
No.~1999/05404-3; Minist\'erio da Ci\^encia, Tecnologia, Inova\c{c}\~oes e
Comunica\c{c}\~oes (MCTIC); Czech Republic -- Grant No.~MSMT CR LTT18004,
LM2015038, LM2018102, CZ.02.1.01/0.0/0.0/16{\textunderscore}013/0001402,
CZ.02.1.01/0.0/0.0/18{\textunderscore}046/0016010 and
CZ.02.1.01/0.0/0.0/17{\textunderscore}049/0008422; France -- Centre de Calcul
IN2P3/CNRS; Centre National de la Recherche Scientifique (CNRS); Conseil
R\'egional Ile-de-France; D\'epartement Physique Nucl\'eaire et Corpusculaire
(PNC-IN2P3/CNRS); D\'epartement Sciences de l'Univers (SDU-INSU/CNRS);
Institut Lagrange de Paris (ILP) Grant No.~LABEX ANR-10-LABX-63 within
the Investissements d'Avenir Programme Grant No.~ANR-11-IDEX-0004-02;
Germany -- Bundesministerium f\"ur Bildung und Forschung (BMBF); Deutsche
Forschungsgemeinschaft (DFG); Finanzministerium Baden-W\"urttemberg;
Helmholtz Alliance for Astroparticle Physics (HAP);
Helmholtz-Gemeinschaft Deutscher Forschungszentren (HGF); Ministerium
f\"ur Innovation, Wissenschaft und Forschung des Landes
Nordrhein-Westfalen; Ministerium f\"ur Wissenschaft, Forschung und Kunst
des Landes Baden-W\"urttemberg; Italy -- Istituto Nazionale di Fisica
Nucleare (INFN); Istituto Nazionale di Astrofisica (INAF); Ministero
dell'Istruzione, dell'Universit\'a e della Ricerca (MIUR); CETEMPS Center
of Excellence; Ministero degli Affari Esteri (MAE); M\'exico -- Consejo
Nacional de Ciencia y Tecnolog\'\i{}a (CONACYT) No.~167733; Universidad
Nacional Aut\'onoma de M\'exico (UNAM); PAPIIT DGAPA-UNAM; The Netherlands
-- Ministry of Education, Culture and Science; Netherlands Organisation
for Scientific Research (NWO); Dutch national e-infrastructure with the
support of SURF Cooperative; Poland -Ministry of Science and Higher
Education, grant No.~DIR/WK/2018/11; National Science Centre, Grants
No.~2013/08/M/ST9/00322, No.~2016/23/B/ST9/01635 and No.~HARMONIA
5--2013/10/M/ST9/00062, UMO-2016/22/M/ST9/00198; Portugal -- Portuguese
national funds and FEDER funds within Programa Operacional Factores de
Competitividade through Funda\c{c}\~ao para a Ci\^encia e a Tecnologia
(COMPETE); Romania -- Romanian Ministry of Education and Research, the
Program Nucleu within MCI (PN19150201/16N/2019 and PN19060102) and
project PN-III-P1-1.2-PCCDI-2017-0839/19PCCDI/2018 within PNCDI III;
Slovenia -- Slovenian Research Agency, grants P1-0031, P1-0385, I0-0033,
N1-0111; Spain -- Ministerio de Econom\'\i{}a, Industria y Competitividad
(FPA2017-85114-P and FPA2017-85197-P), Xunta de Galicia (ED431C
2017/07), Junta de Andaluc\'\i{}a (SOMM17/6104/UGR), Feder Funds, RENATA Red
Nacional Tem\'atica de Astropart\'\i{}culas (FPA2015-68783-REDT) and Mar\'\i{}a de
Maeztu Unit of Excellence (MDM-2016-0692); USA -- Department of Energy,
Contracts No.~DE-AC02-07CH11359, No.~DE-FR02-04ER41300,
No.~DE-FG02-99ER41107 and No.~DE-SC0011689; National Science Foundation,
Grant No.~0450696; The Grainger Foundation; Marie Curie-IRSES/EPLANET;
European Particle Physics Latin American Network; and UNESCO.
\end{sloppypar}

%% file: latex_authorlist_authors.tex

\begin{sloppypar}
A.~Aab$^{75}$,
P.~Abreu$^{67}$,
M.~Aglietta$^{50,49}$,
J.M.~Albury$^{12}$,
I.~Allekotte$^{1}$,
A.~Almela$^{8,11}$,
J.~Alvarez Castillo$^{63}$,
J.~Alvarez-Mu\~niz$^{74}$,
R.~Alves Batista$^{75}$,
G.A.~Anastasi$^{58,49}$
% ...
\end{sloppypar}

%% file: latex_authorlist_institutions.tex
\begin{description}[labelsep=0.2em,align=right,labelwidth=0.7em,labelindent=0em,leftmargin=2em,noitemsep]
\item[$^{1}$] Centro At\'omico Bariloche and Instituto Balseiro (CNEA-UNCuyo-CONICET), San Carlos de Bariloche, Argentina
\item[$^{2}$] Centro de Investigaciones en L\'aseres y Aplicaciones, CITEDEF and CONICET, Villa Martelli, Argentina
\item[$^{3}$] Departamento de F\'\i{}sica and Departamento de Ciencias de la Atm\'osfera y los Oc\'eanos, FCEyN, Universidad de Buenos Aires and CONICET, Buenos Aires, Argentina
\item[$^{4}$] IFLP, Universidad Nacional de La Plata and CONICET, La Plata, Argentina
\item[$^{5}$] Instituto de Astronom\'\i{}a y F\'\i{}sica del Espacio (IAFE, CONICET-UBA), Buenos Aires, Argentina
\item[$^{6}$] Instituto de F\'\i{}sica de Rosario (IFIR) -- CONICET/U.N.R.\ and Facultad de Ciencias Bioqu\'\i{}micas y Farmac\'euticas U.N.R., Rosario, Argentina
\item[$^{7}$] Instituto de Tecnolog\'\i{}as en Detecci\'on y Astropart\'\i{}culas (CNEA, CONICET, UNSAM), and Universidad Tecnol\'ogica Nacional -- Facultad Regional Mendoza (CONICET/CNEA), Mendoza, Argentina
\item[$^{8}$] Instituto de Tecnolog\'\i{}as en Detecci\'on y Astropart\'\i{}culas (CNEA, CONICET, UNSAM), Buenos Aires, Argentina
\item[$^{9}$] Observatorio Pierre Auger, Malarg\"ue, Argentina
\item[$^{10}$] Observatorio Pierre Auger and Comisi\'on Nacional de Energ\'\i{}a At\'omica, Malarg\"ue, Argentina
\item[$^{11}$] Universidad Tecnol\'ogica Nacional -- Facultad Regional Buenos Aires, Buenos Aires, Argentina
\item[$^{12}$] University of Adelaide, Adelaide, S.A., Australia
\item[$^{13}$] Universit\'e Libre de Bruxelles (ULB), Brussels, Belgium
\item[$^{14}$] Vrije Universiteit Brussels, Brussels, Belgium
\item[$^{15}$] Centro Brasileiro de Pesquisas Fisicas, Rio de Janeiro, RJ, Brazil
\item[$^{16}$] Centro Federal de Educa\c{c}\~ao Tecnol\'ogica Celso Suckow da Fonseca, Nova Friburgo, Brazil
\item[$^{17}$] Universidade de S\~ao Paulo, Escola de Engenharia de Lorena, Lorena, SP, Brazil
\item[$^{18}$] Universidade de S\~ao Paulo, Instituto de F\'\i{}sica de S\~ao Carlos, S\~ao Carlos, SP, Brazil
\item[$^{19}$] Universidade de S\~ao Paulo, Instituto de F\'\i{}sica, S\~ao Paulo, SP, Brazil
\item[$^{20}$] Universidade Estadual de Campinas, IFGW, Campinas, SP, Brazil
\item[$^{21}$] Universidade Estadual de Feira de Santana, Feira de Santana, Brazil
\item[$^{22}$] Universidade Federal do ABC, Santo Andr\'e, SP, Brazil
\item[$^{23}$] Universidade Federal do Paran\'a, Setor Palotina, Palotina, Brazil
\item[$^{24}$] Universidade Federal do Rio de Janeiro, Instituto de F\'\i{}sica, Rio de Janeiro, RJ, Brazil
\item[$^{25}$] Universidade Federal do Rio de Janeiro (UFRJ), Observat\'orio do Valongo, Rio de Janeiro, RJ, Brazil
\item[$^{26}$] Universidade Federal Fluminense, EEIMVR, Volta Redonda, RJ, Brazil
\item[$^{27}$] Universidad de Medell\'\i{}n, Medell\'\i{}n, Colombia
\item[$^{28}$] Universidad Industrial de Santander, Bucaramanga, Colombia
\item[$^{29}$] Charles University, Faculty of Mathematics and Physics, Institute of Particle and Nuclear Physics, Prague, Czech Republic
\item[$^{30}$] Institute of Physics of the Czech Academy of Sciences, Prague, Czech Republic
\item[$^{31}$] Palacky University, RCPTM, Olomouc, Czech Republic
\item[$^{32}$] Universit\'e Paris-Saclay, CNRS/IN2P3, IJCLab, Orsay, France
\item[$^{33}$] Laboratoire de Physique Nucl\'eaire et de Hautes Energies (LPNHE), Universit\'es Paris 6 et Paris 7, CNRS-IN2P3, Paris, France
\item[$^{34}$] Univ.\ Grenoble Alpes, CNRS, Grenoble Institute of Engineering Univ.\ Grenoble Alpes, LPSC-IN2P3, 38000 Grenoble, France
\item[$^{35}$] Bergische Universit\"at Wuppertal, Department of Physics, Wuppertal, Germany
\item[$^{36}$] Karlsruhe Institute of Technology, Institute for Experimental Particle Physics (ETP), Karlsruhe, Germany
\item[$^{37}$] Karlsruhe Institute of Technology, Institut f\"ur Kernphysik, Karlsruhe, Germany
\item[$^{38}$] Karlsruhe Institute of Technology, Institut f\"ur Prozessdatenverarbeitung und Elektronik, Karlsruhe, Germany
\item[$^{39}$] RWTH Aachen University, III.\ Physikalisches Institut A, Aachen, Germany
\item[$^{40}$] Universit\"at Hamburg, II.\ Institut f\"ur Theoretische Physik, Hamburg, Germany
\item[$^{41}$] Universit\"at Siegen, Fachbereich 7 Physik -- Experimentelle Teilchenphysik, Siegen, Germany
\item[$^{42}$] Gran Sasso Science Institute, L'Aquila, Italy
\item[$^{43}$] INFN Laboratori Nazionali del Gran Sasso, Assergi (L'Aquila), Italy
\item[$^{44}$] INFN, Sezione di Catania, Catania, Italy
\item[$^{45}$] INFN, Sezione di Lecce, Lecce, Italy
\item[$^{46}$] INFN, Sezione di Milano, Milano, Italy
\item[$^{47}$] INFN, Sezione di Napoli, Napoli, Italy
\item[$^{48}$] INFN, Sezione di Roma ``Tor Vergata'', Roma, Italy
\item[$^{49}$] INFN, Sezione di Torino, Torino, Italy
\item[$^{50}$] Osservatorio Astrofisico di Torino (INAF), Torino, Italy
\item[$^{51}$] Politecnico di Milano, Dipartimento di Scienze e Tecnologie Aerospaziali , Milano, Italy
\item[$^{52}$] Universit\`a del Salento, Dipartimento di Matematica e Fisica ``E.\ De Giorgi'', Lecce, Italy
\item[$^{53}$] Universit\`a dell'Aquila, Dipartimento di Scienze Fisiche e Chimiche, L'Aquila, Italy
\item[$^{54}$] Universit\`a di Catania, Dipartimento di Fisica e Astronomia, Catania, Italy
\item[$^{55}$] Universit\`a di Milano, Dipartimento di Fisica, Milano, Italy
\item[$^{56}$] Universit\`a di Napoli ``Federico II'', Dipartimento di Fisica ``Ettore Pancini'', Napoli, Italy
\item[$^{57}$] Universit\`a di Roma ``Tor Vergata'', Dipartimento di Fisica, Roma, Italy
\item[$^{58}$] Universit\`a Torino, Dipartimento di Fisica, Torino, Italy
\item[$^{59}$] Benem\'erita Universidad Aut\'onoma de Puebla, Puebla, M\'exico
\item[$^{60}$] Centro de Investigaci\'on y de Estudios Avanzados del IPN (CINVESTAV), M\'exico, D.F., M\'exico
\item[$^{61}$] Unidad Profesional Interdisciplinaria en Ingenier\'\i{}a y Tecnolog\'\i{}as Avanzadas del Instituto Polit\'ecnico Nacional (UPIITA-IPN), M\'exico, D.F., M\'exico
\item[$^{62}$] Universidad Aut\'onoma de Chiapas, Tuxtla Guti\'errez, Chiapas, M\'exico
\item[$^{63}$] Universidad Nacional Aut\'onoma de M\'exico, M\'exico, D.F., M\'exico
\item[$^{64}$] Institute of Nuclear Physics PAN, Krakow, Poland
\item[$^{65}$] University of \L{}\'od\'z, Faculty of Astrophysics, \L{}\'od\'z, Poland
\item[$^{66}$] University of \L{}\'od\'z, Faculty of High-Energy Astrophysics,\L{}\'od\'z, Poland
\item[$^{67}$] Laborat\'orio de Instrumenta\c{c}\~ao e F\'\i{}sica Experimental de Part\'\i{}culas -- LIP and Instituto Superior T\'ecnico -- IST, Universidade de Lisboa -- UL, Lisboa, Portugal
\item[$^{68}$] ``Horia Hulubei'' National Institute for Physics and Nuclear Engineering, Bucharest-Magurele, Romania
\item[$^{69}$] Institute of Space Science, Bucharest-Magurele, Romania
\item[$^{70}$] University Politehnica of Bucharest, Bucharest, Romania
\item[$^{71}$] Center for Astrophysics and Cosmology (CAC), University of Nova Gorica, Nova Gorica, Slovenia
\item[$^{72}$] Experimental Particle Physics Department, J.\ Stefan Institute, Ljubljana, Slovenia
\item[$^{73}$] Universidad de Granada and C.A.F.P.E., Granada, Spain
\item[$^{74}$] Instituto Galego de F\'\i{}sica de Altas Enerx\'\i{}as (IGFAE), Universidade de Santiago de Compostela, Santiago de Compostela, Spain
\item[$^{75}$] IMAPP, Radboud University Nijmegen, Nijmegen, The Netherlands
\item[$^{76}$] KVI -- Center for Advanced Radiation Technology, University of Groningen, Groningen, The Netherlands
\item[$^{77}$] Nationaal Instituut voor Kernfysica en Hoge Energie Fysica (NIKHEF), Science Park, Amsterdam, The Netherlands
\item[$^{78}$] Stichting Astronomisch Onderzoek in Nederland (ASTRON), Dwingeloo, The Netherlands
\item[$^{79}$] Universiteit van Amsterdam, Faculty of Science, Amsterdam, The Netherlands
\item[$^{80}$] Case Western Reserve University, Cleveland, OH, USA
\item[$^{81}$] Colorado School of Mines, Golden, CO, USA
\item[$^{82}$] Department of Physics and Astronomy, Lehman College, City University of New York, Bronx, NY, USA
\item[$^{83}$] Louisiana State University, Baton Rouge, LA, USA
\item[$^{84}$] Michigan Technological University, Houghton, MI, USA
\item[$^{85}$] New York University, New York, NY, USA
\item[$^{86}$] Pennsylvania State University, University Park, PA, USA
\item[$^{87}$] University of Chicago, Enrico Fermi Institute, Chicago, IL, USA
\item[$^{88}$] University of Delaware, Department of Physics and Astronomy, Bartol Research Institute, Newark, DE, USA
\item[] -----
\item[$^{a}$] School of Physics and Astronomy, University of Leeds, Leeds, United Kingdom
\item[$^{b}$] Max-Planck-Institut f\"ur Radioastronomie, Bonn, Germany
\item[$^{c}$] Fermi National Accelerator Laboratory, USA
\item[$^{d}$] also at Universidade Federal de Alfenas, Po\c{c}os de Caldas, Brazil
\item[$^{e}$] Colorado State University, Fort Collins, CO, USA
\item[$^{f}$] now at Hakubi Center for Advanced Research and Graduate School of Science, Kyoto University, Kyoto, Japan
\item[$^{g}$] also at Karlsruhe Institute of Technology, Karlsruhe, Germany
\item[$^{h}$] also at Radboud Universtiy Nijmegen, Nijmegen, The Netherlands
\item[$^{i}$] also at University of Bucharest, Physics Department, Bucharest, Romania
\end{description}